\documentclass[modern, letterpaper]{aastex62}

\usepackage{amsmath}
\usepackage{amsfonts}
\usepackage{amssymb}
\usepackage{booktabs}

\usepackage{graphicx}
\usepackage{color}

\definecolor{cbblue}{HTML}{3182bd}
\usepackage{hyperref}
\definecolor{linkcolor}{rgb}{0.02,0.35,0.55}
\definecolor{citecolor}{rgb}{0.45,0.45,0.45}
\hypersetup{colorlinks=true,linkcolor=linkcolor,citecolor=citecolor,
            filecolor=linkcolor,urlcolor=linkcolor}
\hypersetup{pageanchor=true}

\newcommand{\documentname}{\textsl{Article}}
\newcommand{\sectionname}{Section}
\renewcommand{\figurename}{Figure}
\newcommand{\eqname}{Equation}
\renewcommand{\tablename}{Table}

\newcommand{\package}[1]{\textsl{#1}}
\newcommand{\acronym}[1]{{\small{#1}}}

\newcommand{\python}{\package{Python}}

\newcommand{\project}[1]{\textsl{#1}}

\newcommand{\given}{\,|\,}

\newcommand{\dd}{\mathrm{d}}

\newcommand{\msun}{\ensuremath{\mathrm{M}_\odot}}
\newcommand{\kms}{\ensuremath{\mathrm{km}~\mathrm{s}^{-1}}}
\newcommand{\mps}{\ensuremath{\mathrm{m}~\mathrm{s}^{-1}}}

\newcommand{\bs}[1]{\boldsymbol{#1}}
\definecolor{mahogany}{RGB}{165,15,21}

\graphicspath{{}}

\newcommand{\apogee}{\project{\acronym{APOGEE}}}
\newcommand{\sdssiv}{\project{\acronym{SDSS-IV}}}
\newcommand{\thejoker}{\project{The~Joker}}
\newcommand{\thecannon}{\project{The~Cannon}}
\newcommand{\DR}{\acronym{DR14}}
\newcommand{\DRtw}{\acronym{DR12}}
\newcommand{\RC}{\acronym{RC}}
\newcommand{\RGB}{\acronym{RGB}}

\newcommand{\pdf}{\textit{pdf}}
\newcommand{\logg}{\ensuremath{\log g}}
\newcommand{\Teff}{\ensuremath{T_{\textrm{eff}}}}

\newcommand{\nprior}{536,870,912}
\newcommand{\nposterior}{256}
\newcommand{\nstars}{96,231}
\newcommand{\nvisits}{397,559}
\newcommand{\ncontrol}{16,384}
\newcommand{\lnKcut}{-0.2}
\newcommand{\nhighK}{4,898}
\newcommand{\nbimodal}{106}
\newcommand{\nunimodal}{320}

\shortauthors{Price-Whelan et al.}

\begin{document}\sloppy\sloppypar\raggedbottom\frenchspacing 

\title{Binary companions of evolved stars in \apogee\ \DR: \\
       Search method and catalog of $\sim$5,000 companions}

\author[0000-0003-0872-7098]{Adrian~M.~Price-Whelan}
\affiliation{Department of Astrophysical Sciences,
             Princeton University, Princeton, NJ 08544, USA}
\email{adrn@astro.princeton.edu}
\correspondingauthor{Adrian M. Price-Whelan}

\author[0000-0003-2866-9403]{David~W.~Hogg}
\affiliation{Max-Planck-Institut f\"ur Astronomie,
             K\"onigstuhl 17, D-69117 Heidelberg, Germany}
\affiliation{Center for Cosmology and Particle Physics,
             Department of Physics,
             New York University, 726 Broadway,
             New York, NY 10003, USA}
\affiliation{Center for Data Science,
             New York University, 60 Fifth Ave,
             New York, NY 10011, USA}
\affiliation{Flatiron Institute,
             Simons Foundation,
             162 Fifth Avenue,
             New York, NY 10010, USA}

\author[0000-0003-4996-9069]{Hans-Walter~Rix}
\affiliation{Max-Planck-Institut f\"ur Astronomie,
             K\"onigstuhl 17, D-69117 Heidelberg, Germany}

\author{Nathan De Lee}
\affiliation{Department of Physics, Geology, and Engineering Technology, Northern Kentucky Univerity, Highland Heights, KY 41099}
\affiliation{Department of Physics \& Astronomy, Vanderbilt University, Nashville, TN 37235}

\author{Steven R. Majewski}
\affiliation{Department of Astronomy, University of Virginia, Charlottesville, VA 22904-4325, USA}

\author{David L. Nidever}
\affiliation{Department of Physics, Montana State University, P.O. Box
173840, Bozeman, MT 59717-3840}
\affiliation{National Optical Astronomy Observatory, 950 North Cherry
Ave, Tucson, AZ 85719}

\author{Nicholas Troup}
\affiliation{Department of Physics, Salisbury University, Salisbury, MD 21801}

\author{Jos\'e G. Fern\'andez-Trincado}
\affiliation{Departamento de Astronom\'ia, Universidad de Concepci\'on, Casilla 160-C, Concepci\'on, Chile}

\author{Domingo A. Garc\'ia-Hern\'andez}
\affiliation{Instituto de Astrof\'isica de Canarias (IAC), E-38205 La Laguna, Tenerife, Spain}
\affiliation{Universidad de La Laguna (ULL), Departamento de Astrof\'isica, E-38206 La Laguna, Tenerife, Spain}

\author{Pen\'elope Longa-Pe\~na}
\affiliation{Unidad de Astronom\'ia, Universidad de Antofagasta, Avenida Angamos 601, Antofagasta 1270300, Chile}

\author{Christian Nitschelm}
\affiliation{Unidad de Astronom\'ia, Universidad de Antofagasta, Avenida Angamos 601, Antofagasta 1270300, Chile}

\author{Jennifer Sobeck}
\affiliation{Department of Astronomy, University of Washington, Box 351580,
Seattle, WA 98195, USA}

\author{Olga Zamora}
\affiliation{Instituto de Astrof\'isica de Canarias (IAC), E-38205 La Laguna, Tenerife, Spain}
\affiliation{Universidad de La Laguna (ULL), Departamento de Astrof\'isica, E-38206 La Laguna, Tenerife, Spain}

\begin{abstract}\noindent 
Multi-epoch radial velocity measurements of stars can be used to identify
stellar, sub-stellar, and planetary-mass companions.
Even a small number of observation epochs can be informative about
companions, though there can be multiple qualitatively different orbital
solutions that fit the data.
We have custom-built a Monte Carlo sampler (\thejoker) that delivers reliable
(and often highly multi-modal) posterior samplings for companion orbital
parameters given sparse radial-velocity data.
Here we use \thejoker\ to perform a search for companions to \nstars\ red-giant
stars observed in the \apogee\ survey (\DR) with $\geq 3$ spectroscopic epochs.
We select stars with probable companions by making a cut on our posterior belief
about the amplitude of the stellar radial-velocity variation induced by the
orbit.
We provide (1)~a catalog of \nunimodal\ companions for which the stellar
companion properties can be confidently determined, (2)~a catalog of \nhighK\
stars that likely have companions, but would require more observations to
uniquely determine the orbital properties, and (3)~posterior samplings for the
full orbital parameters for all stars in the parent sample.
We show the characteristics of systems with confidently determined companion
properties and highlight interesting systems with candidate compact object
companions.
\end{abstract}


\section{Introduction} \label{sec:intro}

Time-domain radial-velocity measurements of stars contain information about
massive companions.
Even with two successive observations of a single star, a difference in the
measured radial velocities implies the existence of at least one companion.
However, with few or imprecise radial-velocity measurements, the orbital
properties of the companion(s) may be poorly constrained (e.g.,
\citealt{Price-Whelan:2017}).
The vast majority of spectroscopic targets with repeat observations in the
largest (by number of objects) stellar spectroscopic surveys are observed just a
few times with sparse, non-uniform phase coverage.
Most prior searches for companions using survey RV data have therefore
restricted their searches to only sources with many, high-precision epochs, so
that the orbital solution can be unambiguously determined (e.g.,
\citealt{Troup:2016}), or have used simple statistics computed from the data to
study multiplicity (e.g., $\textrm{RV}_\textrm{max}$; \citealt{Badenes:2017}).

If there are only a few radial-velocity epochs per star, and if the companion
spectrum is not observed, the data will be consistent with many different
combinations of primary orbital parameters (period, amplitude, eccentricity,
etc.).
To identify companions to the typical star observed in a spectroscopic survey,
we therefore face at least one major challenge: how, given a small number of
observations of a primary star, do we reliably obtain posterior information
about the binary-system properties?
In general, the likelihood function---and the posterior probability distribution
function (\pdf) under any reasonable prior \pdf---will be highly multimodal, and
many of the modes will have comparable integrated probability density.
For example, with just two radial-velocity measurements, a harmonic series of
period modes will exist in the likelihood function.

We have solved the problem of deriving comprehensive multi-modal \pdf\ samplings
previously with \thejoker\ (\citealt{Price-Whelan:2017}), a Monte Carlo
rejection sampler that is computationally expensive but probabilistically
righteous:
It delivers independent posterior \pdf\ samples for single-companion binary
orbital parameters, given any number of radial-velocity measurements.
Here we use \thejoker\ to generate posterior \pdf\ samples for stars observed by
the \apogee\ survey (see \sectionname~\ref{sec:data}; \citealt{Majewski:2017}).

The \apogee\ surveys primarily target red-giant-branch (\RGB) and other evolved
stars (e.g., red clump giants, RC), which are ideal for the study of single-line
binary systems.
In general, they are unlikely to have comparably-bright companions, and their
spectra are therefore fit as single-line objects.
When this constraint is not met, \thejoker\ will in general fail, and a model
that fits for a mixture of stellar spectra is more appropriate (e.g.,
\citealt{El-Badry:2018}).
The subset of RC stars are even more powerful as they are standard candles and
have masses that can be estimated using spectroscopy (using dredged-up elements;
\citealt{Martig:2016,Ness:2016}).
With mass estimates for the primary star, the binary-orbit fitting will return
$M_2\,\sin i$ (minimum mass) estimates for the secondary, and not just estimates
of the so-called ``binary mass function.''
Additionally, the \apogee\ pipelines (\citealt{Garcia-Perez:2016}) and also
\thecannon\ (\citealt{Ness:2015}) produce detailed abundance estimates for \RGB\
and \RC\ stars.
If there are causal relationships between chemical abundances and binary
companions---as are expected---these should be measurable.

By making cuts on this library of posterior \pdf\ samples (described in detail
in \sectionname~\ref{sec:catalogs}), we deliver catalogs of binary-star systems
from the \apogee\ survey, and show the bulk properties of these systems.

Binary and multiple star systems are of great interest in astrophysics: The
population of stars and their companions encodes information about star
formation processes, stellar parameters and evolution, and the dynamics of
multi-body systems (for recent reviews, see \citealt{Duchene:2013,Moe:2017}).
Most of what is known about stellar companions comes from studies of nearby
main-sequence (MS) stars (e.g.,
\citealt{Duquennoy:1991,Raghavan:2010,Tokovinin:2014,Moe:2017}).
Nearly 50\% of MS stars in the solar neighborhood have companions
(e.g., \citealt{Tokovinin:2014}).
MS stars with companions have a large dynamic range of constituent and orbital
characteristics.
For example, binary stars have mass-ratios that span from $\approx 0.03$
to 1 (e.g., \citealt{Kraus:2008}), and have periods from days to
millions of years (e.g., \citealt{Raghavan:2010}).
Less is known about population properties of non-interacting or detached
companions to evolved stars.
The catalogs and methodology presented in this work are a first step towards
performing population inferences of binary star systems with evolved star
members.

\section{Data} \label{sec:data}

All data used in this work come from the publicly-available data release 14
(\DR) of the \apogee\ survey (\citealt{Majewski:2017,Abolfathi:2017}), a
component of the Sloan Digital Sky Survey IV (\sdssiv;
\citealt{Gunn:2006,Blanton:2017}).
\apogee\ is designed to map stars across much of the Milky Way by obtaining
high-resolution ($R \sim 22,500$) infrared ($H$-band) spectroscopy of primarily
\RGB\ stars.
Targets are selected with simple color and brightness cuts, but the survey uses
fiber-plugged plates with a maximum of 300 fibers per each $\approx
1.5~\textrm{deg}^2$ field of view, leading to ``pencil-beam''-like sampling of
the Galactic stellar distribution.
In order to meet signal-to-noise ratio requirements, most \apogee\ stars are
observed multiple times in a series of ``visits,'' typically with at least one
visit separated by a month or more in order to help identify binary stars.

Data taken as part of the \apogee\ survey are reduced with a multi-step data
reduction pipeline that ultimately solves for the stellar parameters, chemical
abundances, and radial velocities for each target
(\citealt{Nidever:2015}).
Most relevant for this work, the visit radial velocities (RVs) are determined
using an iterative scheme: the individual visit spectra are combined using
initial guesses for the relative RVs into a coadded spectrum, which is then used
to re-derive the relative visit velocities.
The stellar parameters---surface gravity, \logg, and effective temperature,
\Teff---and the chemical abundances are determined from the coadded
spectrum as a part of the \apogee\ Stellar Parameters and Chemical Abundances
Pipeline (\acronym{ASPCAP}; \citealt{ASPCAP}).

\begin{figure}[h]
\begin{center}
\includegraphics[width=0.7\textwidth]{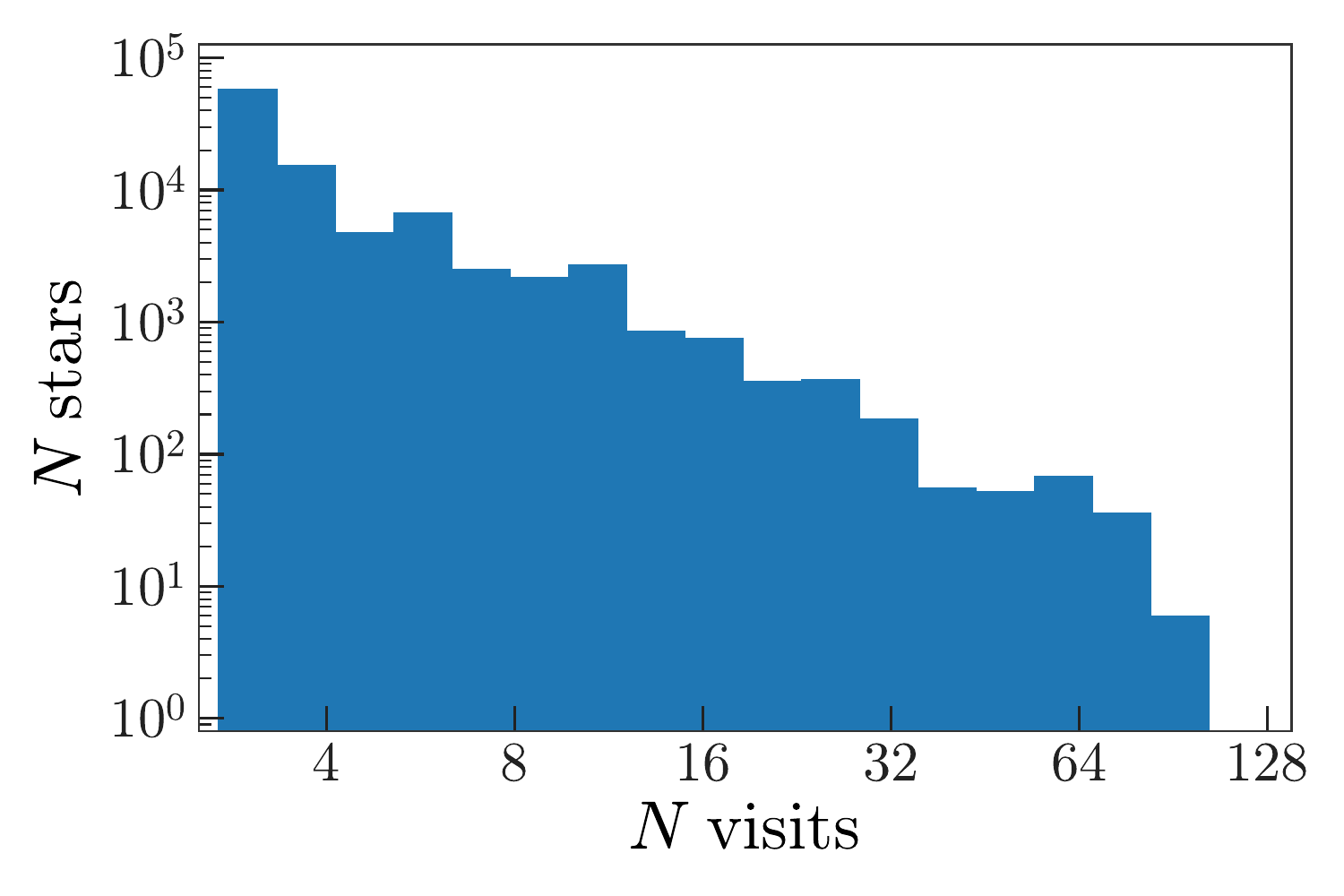}
\end{center}
\caption{%
Number of \apogee\ \DR\ stars in logarithmic bins of number of visits that pass
the quality cuts described in \sectionname~\ref{sec:data}.
In total, this work uses \nvisits\ visits and \nstars\ unique sources.
\label{fig:nvisits}
}
\end{figure}

\begin{figure}[h]
\begin{center}
\includegraphics[width=\textwidth]{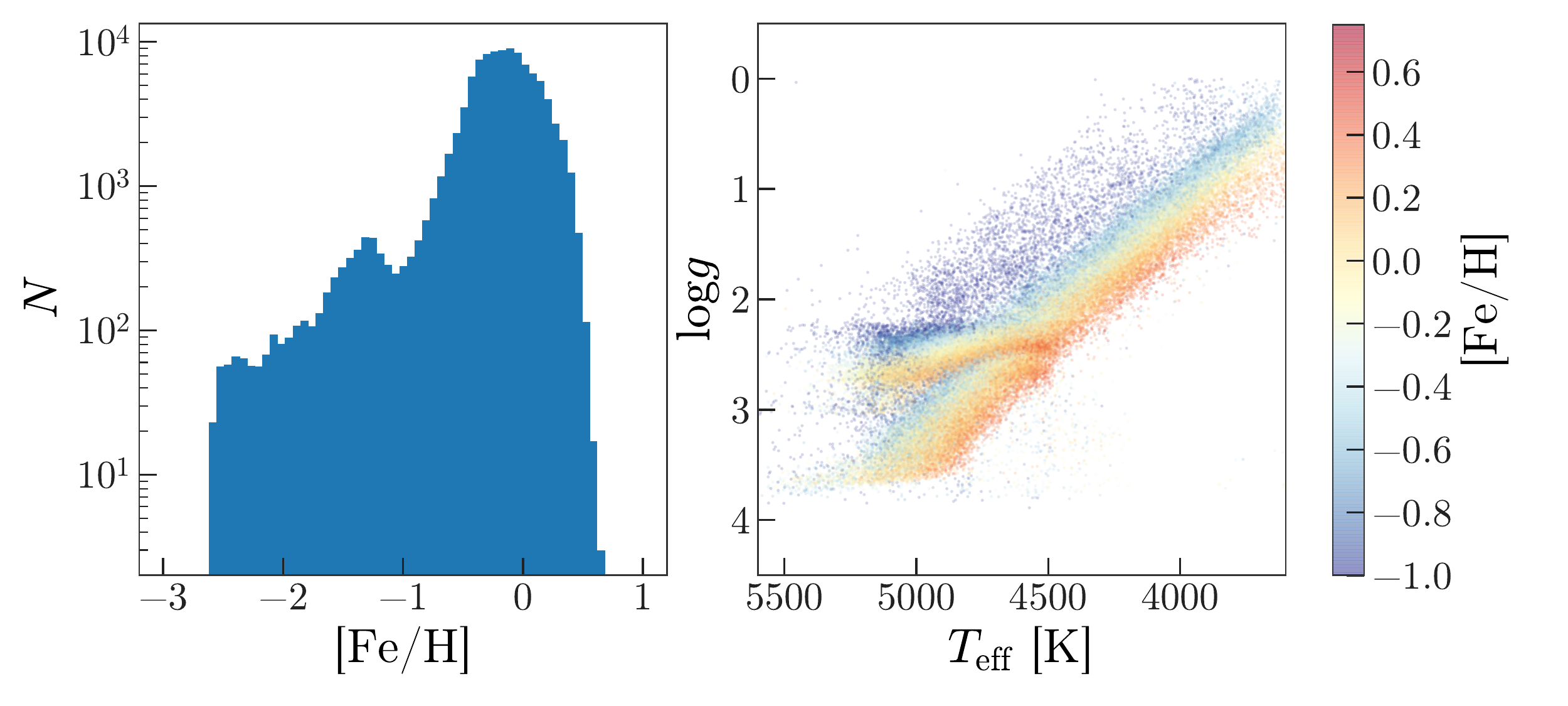}
\end{center}
\caption{%
\textit{Left panel}: Number of stars in bins of iron abundance,
$[\textrm{Fe}/\textrm{H}]$, that pass the quality cuts described in
\sectionname~\ref{sec:data}.
\textit{Right panel}: Distribution of stars in our sample in stellar parameters
log-surface-gravity, \logg, and effective temperature, \Teff, with points
colored by the iron abundance.
\label{fig:loggteff}
}
\end{figure}

We use the primary data products from \apogee\ \DR\ (i.e. the \texttt{allStar}
and \texttt{allVisit} files) which contain 258,475 unique source IDs
(\texttt{APOGEE\_ID}) and 1,054,381 unique visits.
We select all stars with $\geq 3$ visits that each pass a set of quality cuts,
described below.
For each visit, we require that the visit velocity uncertainty is $< 100~\kms$
(\texttt{VRELERR}) and the following bits are not set in the \texttt{STARFLAGS}
bitmask: \texttt{PERSIST\_HIGH}, \texttt{PERSIST\_JUMP\_POS},
\texttt{PERSIST\_JUMP\_NEG}, \texttt{VERY\_BRIGHT\_NEIGHBOR}, \texttt{LOW\_SNR}.
For each star, we require that $0 < \logg < 4$ and the following bits are not
set in the \texttt{ASPCAPFLAGS} bitmask: \texttt{STAR\_BAD}.
After these cuts, and the requirement of $\geq 3$ visits for a given star, we
are left with \nvisits\ visits for \nstars\ unique sources.
\figurename~\ref{fig:nvisits} shows the number of stars in several bins of
number of visits that pass the above quality cuts: 92\% of the stars in
our parent sample have $< 8$ visits.
\figurename~\ref{fig:loggteff} shows the stellar parameters of all \nstars\
stars in the parent sample used in this work.

\section{Methods}

\subsection{Orbit inference and velocity modeling}
\label{sec:fitting}

For every source in the sample of \apogee\ stars defined in
\sectionname~\ref{sec:data}, we obtain a posterior sampling in binary-system
parameter space, treating it as a single-lined (SB1) spectroscopic binary system
with a single companion.
This sampling is performed with \thejoker\ (\citealt{Price-Whelan:2017}) under a
relatively uninformative prior \pdf, and the resulting posterior samplings are
used to discover and characterize individual binary-star systems and generate a
catalog of companions.
We now describe the assumptions and method used to generate samplings for
individual systems:
\begin{description}
\item[no multiple companions] All radial-velocity variations of the primary
  star are induced by a single companion.
  This is motivated by the idea that triple star systems are usually
  hierarchical so that the period of the inner binary is typically much shorter
  than the orbital period of the outer body (e.g., \citealt{Tokovinin:2018}).
  At present, we ignore the possibility of higher-order multiple systems.
\item[Kepler] All velocity variations of the primary
  are gravitational, and we therefore ignore the possibility of coherent
  intrinsic variation from, e.g., stellar oscillations.
\item[SB1] All spectra are single-lined; that is, we assume that the secondary
  is significantly fainter and is thus undetected in the spectra.
  This assumption is motivated by the fact that we expect \RGB\ stars to be
  substantially more luminous than their typical companion.
  However, there are known main-sequence double-lined binary stars in the
  \apogee\ catalog (\citealt{El-Badry:2018}), and an expected but unknown
  fraction of \RGB--\RGB\ binaries.
\item[simple noise model] Measurements are unbiased and noise estimates are
  correct up to an unknown excess variance.
  All noise contributions result in Gaussian uncertainties on the individual
  radial-velocity measurements.
\end{description}

\thejoker\ is a custom-built Monte Carlo sampler designed to produce independent
posterior samples in Keplerian orbital parameters, given radial-velocity
measurements under the assumptions listed above.
Our parametrization of the orbital elements is similar to the notation in
\citet{Murray:2010}:
The radial velocity $v$ at time $t$ is given by
\begin{equation}
  v(t;\bs{\theta}) = v_0 + K\,[\cos\left(\omega + f(t; e, P, M_0, t_0)\right) +
    e\,\cos\omega]
\end{equation}
where $\bs{\theta} = (P,e,M_0,\omega,K,v_0)$---period, eccentricity, mean
anomaly at a reference time $t_0$, argument of pericenter, velocity
semi-amplitude, barycentric velocity---and the true anomaly, $f$, is a function
of time and the specified parameters (see \sectionname~2 of
\citealt{Price-Whelan:2017} or \eqname~63 in \citealt{Murray:2010}).
In addition to the orbital parameters listed above, \thejoker\ can also generate
samples in an ``excess variance'' parameter, $s^2$, that is added to the
per-visit measurement variances.
This parameter allows us to test whether the visit RV uncertainties are
underestimated
For upper \RGB\ stars the inferred excess variance will be a combination of
extra systematic uncertainty and true astrophysical surface jitter
(e.g., \cite{Hekker:2008}).

\thejoker\ was designed for the extremely multi-modal \pdf 's expected when the
number of radial-velocity measurements of a source is small, or the data are
sparse (in phase-coverage) or noisy.
While other Markov Chain Monte Carlo (MCMC) methods have difficulty producing
independent samples with such data, \thejoker\ succeeds by brute force:
After generating an initial (very large) library of prior samples from an
assumed prior \pdf\ (see below), the (typically multi-modal) likelihood is
evaluated at each sample and used to rejection sample.
In practice, given a number of requested samples for each star, the sampling
proceeds iteratively: since it is easier to accept samples when the data is
sparse or noisy, far more prior sample draws (and thus likelihood evaluations)
must occur under very constraining data.

%
%
%

\subsection{Individual-system posterior samplings}
\label{sec:samplings}

Here we describe the specifics of generating posterior samplings for all of the
\apogee\ targets in our parent sample.
We execute the full procedure twice for different goals (as described in
\sectionname~\ref{sec:catalogs}), and only here outline the key steps in the
pipeline.

For all \nstars\ \apogee\ stars with $\geq 3$ good visits (see
\sectionname~\ref{sec:data}), we use \thejoker\ to generate posterior samplings
for each star under the assumptions listed above (see
\sectionname~\ref{sec:fitting}).
We start by generating a library of \nprior\ prior samples generated under a
prior similar to that defined in \citet{Price-Whelan:2017}:
\begin{itemize}
    \item uniform or isotropic in angle parameters,
    \item uniform in log-period over the domain $[1,32768]~\textrm{day}$,
    \item a beta distribution over eccentricity (using parameters from
    \citealt{Kipping:2013}).
\end{itemize}
For the excess variance parameter, $s^2$, we use a Gaussian over the transformed
parameter $y = \ln s^2$ with the mean and standard deviation $(\mu_y, \sigma_y)$
indicated where the runs are described (see \sectionname~\ref{sec:catalogs}).
The reference time for each star is set to the first visit epoch; $M_0$ then
becomes the mean anomaly at the first visit observation.
\tablename~\ref{tbl:params} contains descriptions of all parameters and priors
used.

\begin{table}[h]
    \centering
    \begin{tabular}{ r l l }
    \hline
    name & prior & description \\
    \hline
    $P$ & $\ln P \sim \mathcal{U}(1, 32768)~\textrm{day}$ & period \\
    $e$ & $e \sim \textrm{Beta}(0.867, 3.03)$ & eccentricity \\
    $t_0$ & fixed & reference time \\
    $M_0$ & $M_0 \sim \mathcal{U}(0, 2\pi)~\textrm{rad}$ & mean anomaly at reference time \\
    $\omega$ & $\omega \sim \mathcal{U}(0, 2\pi)~\textrm{rad}$ & argument of pericenter \\
    $s^2$ & $\ln s^2 \sim \mathcal{N}(\mu_y, \sigma_y^2)$ & extra variance added to each visit variance \\
    $K$ & $\mathcal{N}(0, \sigma_{v}^2)~\kms$ & velocity semi-amplitude \\
    $v_0$ & $\mathcal{N}(0, \sigma_{v}^2)~\kms$ & system barycentric velocity \\
    \hline
    \end{tabular}
    \caption{Summary and description of parameters. $\textrm{Beta}(a, b)$ is the
    beta distribution with shape parameters $(a, b)$, $\mathcal{U}(a, b)$ the
    uniform distribution over the domain $(a, b)$, and $\mathcal{N}(\mu,
    \sigma^2)$ is the normal distribution with mean $\mu$ and variance
    $\sigma^2$.
    For the systemic velocity and semi-amplitude, $(v_0, K)$, we use a broad
    Gaussian prior that is formally inconsistent between \thejoker\ and
    follow-up MCMC sampling (see \sectionname~\ref{sec:mcmc}): in \thejoker\ we
    assume Gaussian priors with $\sigma_v$ much larger than the measurement
    uncertainty so they can be neglected ($\sigma_v \approx \infty$), whereas
    when running MCMC we fix $\sigma_v = 10^3~\kms$.}
    \label{tbl:params}
\end{table}

We request \nposterior\ posterior samples for each source.
Depending on the data quality and phase coverage of the visits, \thejoker\ will
require different numbers of prior samples in order to rejection-sample down to
the requested number of posterior samples: For few-epoch or noisy RV data, many
prior samples will pass the rejection step, whereas for very precise or
many-epoch RV data, \thejoker\ may need to process the full library of prior
samples.
We therefore generate the posterior samples using an iterative procedure that
adaptively predicts how many prior samples to test for each star.
For sources with very constraining data, \thejoker\ may return fewer than the
requested number of samples (as few as one sample).
When this occurs, we continue sampling either using standard MCMC, or by
increasing the size of the prior cache and continuing rejection sampling with
\thejoker.

\subsubsection{``Needs MCMC'': Following up \thejoker\ with MCMC}
\label{sec:mcmc}

If just one posterior sample is returned after exhausting the full library of
prior samples, or if multiple (but fewer than \nposterior) are returned that all
lie within a single mode of the posterior \pdf, the posterior \pdf\ over orbital
parameters is treated as effectively unimodal: these stars are flagged ``needs
MCMC.''
In this case, we use the location of the returned sample (if only one is
returned), or a randomly chosen sample from those returned (if multiple samples
are returned within one mode) to generate a small Gaussian ball of initial
conditions and use standard MCMC to continue sampling until we obtain
\nposterior\ samples.

In detail, we use an ensemble MCMC sampling algorithm (\citealt{Goodman:2010})
implemented in \python\ (\package{emcee}; \citealt{Foreman-Mackey:2013}) to
perform the samplings.
We transform the standard Keplerian orbital parameters to a safer
parametrization, $(\ln P, \sqrt{K}\,\cos M_0, \sqrt{K}\,\sin M_0, \sqrt{e}\,\cos
\omega, \sqrt{e}\,\cos \omega, \ln s^2, v_0)$, for MCMC sampling.
This reparametrization is safer and more efficient for sampling with
\package{emcee}, which expects parameters to be components of a vector so that
linear operations can be applied (see, e.g., \citealt{Hogg:2017}); the angle
variables $(\omega, M_0)$ don't meet this requirement in the standard
parametrization.
We use the same prior \pdf 's as in \thejoker\ when running MCMC (see
\tablename~\ref{tbl:params}).

For each star that is flagged ``needs MCMC,'' we run \package{emcee} with 1024
walkers for 16384 steps, take the final walker positions, and downsample at
random until we have \nposterior\ samples.
We compute the Gelman--Rubin convergence statistic, $\hat{R}_j$,
(\citealt{Gelman:1992}) for each parameter $j$ and include these values in the
catalogs below when standard MCMC is run.
We also provide a binary flag, ``converged,'' for each sampling continued with
MCMC that is set to true if:
\begin{equation}
\underset{j}{\textrm{mean}}\left(\hat{R}_j\right) < 1.1 \quad .
\end{equation}

\subsubsection{``Needs more prior samples'': Extending \thejoker\ sampling}

If more than one posterior sample is returned after exhausting the full library
of prior samples, and the samples lie in multiple modes of the posterior \pdf,
the only way to proceed is to generate more prior samples and continue running
\thejoker: these stars are flagged ``needs more prior samples.''
In this case, we generate another equal-sized library of prior samples (a total
of $2\times\nprior$ samples) and re-do the rejection sampling.
We note that because of the way the rejection sampling step is done, this is not
equivalent to concatenating the results from a second, independent run of
\thejoker: the log-likelihood values for all of the prior samples must be used.
If at the end of this second run the target still has fewer than \nposterior\
samples, the sampling is flagged as ``incomplete.''

\subsection{Null comparison sample}
\label{sec:control-sample}

We construct a comparison sample of simulated data with no RV variability to
assess our false-positive rates in the selections below (see
\sectionname~\ref{sec:catalogs}).
We randomly pick \ncontrol\ stars from the parent sample used in this work and
replace the visit velocity measurements with simulated data.
For each star $n$, we randomly sample an excess variance parameter value from
the prior, $s_n$.
For each visit $k$, we then sample a new velocity $v_{nk}$ by drawing from a
Gaussian with mean equal to the mean of the real data visit velocities,
$\bar{v}_n$, and variance equal to the sum of the visit variance (uncertainty),
$\sigma_{nk}$, and the excess variance,
\begin{equation}
    v_{nk} \sim \mathcal{N}(\bar{v}_n, \sigma_{nk}^2 + s_n^2) \quad .
\end{equation}
When we save the comparison data, we only store the visit uncertainty and
``forget'' the fact that the simulated data is generated with excess variance
(to be inferrred with \thejoker).

%

%

\section{Experiment: infer the excess variance distribution}
\label{sec:inferjitter}

With posterior samplings for all \apogee\ \DR\ systems in hand, and as an
initial use of the per-source posterior samplings, we use a hierarchical
Bayesian model to infer the parameters of the (assumed Gaussian) prior over the
log-excess-variance parameter, $(\mu_y, \sigma_y)$ (see
\tablename~\ref{tbl:params}).
This inference serves as a test-case for future work, where we intend to use the
independent posterior samplings to construct a hierarchical inference over
companion population properties.
This is also a test of the visit velocity uncertainties reported in the \apogee\
data products: If the catalog uncertainties are significantly underestimated, we
expect the inferred log-excess-variance distribution parameters to tend towards
larger values.

In detail, we maximize the marginal likelihood of the population-level
parameters $(\mu_y, \sigma_y)$.
We compute this marginal likelihood using the per-object posterior samples
re-weighted by the ratio of the value of the hyperprior evaluated at a given,
new set of parameters $(\mu_y, \sigma_y)$ over the value of the default prior at
the previously assumed values, $(\mu_{y,0}, \sigma_{y,0})$ (see above).
This trick has been used in other hierarchical inferences as a way to
marginalize over the per-object parameters to infer population-level parameters
(\citealt{Hogg:2010,Foreman-Mackey:2014}); we describe how to compute the
marginal likelihood in detail in Appendix~\ref{sec:hierarch}.

We execute a full run of \thejoker\ on all \nstars\ \apogee\ stars in our parent
sample using initial values for the excess variance distribution parameters
chosen so that $\sqrt{e^{\mu_{y,0}}} \approx 200~\mps$: $(\mu_{y,0},
\sigma_{y,0}) = (10.6, 3)$ in units of $\mps$.
This run took approximately 300 hours on a compute cluster with 448 cores, with
the time dominated by sources with many ($\gtrsim 10$) visits.  For this initial
run, we do not follow up on stars that return $<$\nposterior\ samples.

\begin{figure}[h]
\begin{center}
\includegraphics[width=\textwidth]{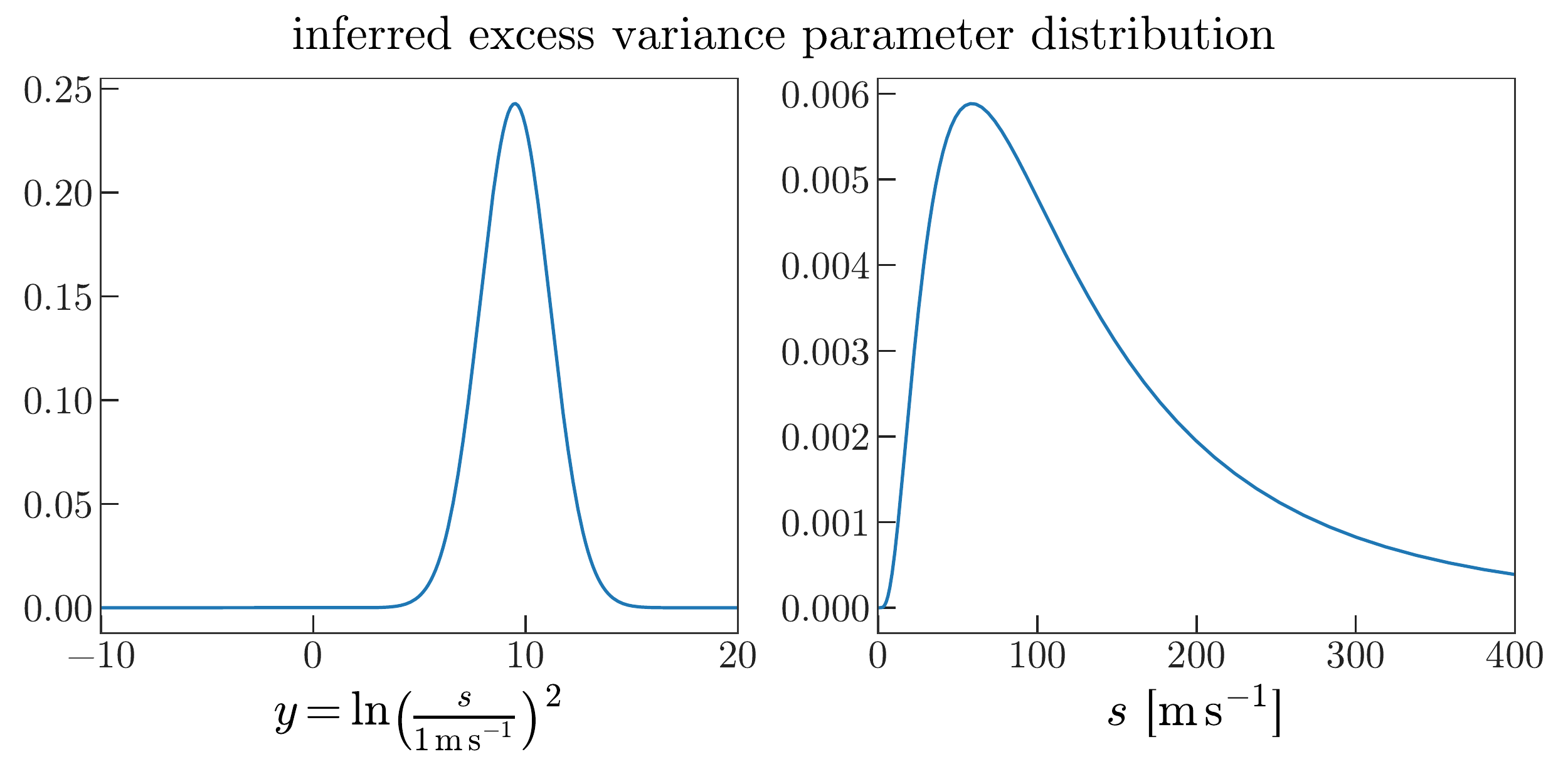}
\end{center}
\caption{%
Inferred prior over excess variance parameter ($y = \ln s^2$) using posterior
samples for 1,825 lower-RGB stars with $>10$ visits.
Left and right panels show the distribution corresponding to the
maximum-likelihood parameters $(\mu_y, \sigma_y) = (9.50, 1.64)$ in log and
linear, respectively.
\label{fig:infer-jitter}
}
\end{figure}

We use 1,825 stars with $>10$ visits and $\logg > 2$ (to avoid upper RGB stars
that have large intrinsic jitter) and maximize the above likelihood to determine
better hyperparameters for the log-excess-variance parameter distribution.
\figurename~\ref{fig:infer-jitter} shows the distribution corresponding to the
maximum-likelihood hyperparameters, $\alpha^* = (\mu_y^*, \sigma_y^*) = (9.50,
1.64)$.
These values are consistent with the estimated systematic floor of the visit
velocity uncertainties of $\approx 100$--$200~\mps$ estimated from stars
observed multiple times and on multiple plates (\citealt{Nidever:2015}).

However, there are a number of caveats to keep in mind about this estimate of
the excess variance distribution.
First, we do not remove triple or other multiple systems: For any individual
system, the radial velocity variations induced by other massive bodies will lead
to larger preferred values for the excess variance parameter.
We do not expect there to be a large number of triple systems in our sample, but
this is still an important consideration for future efforts.
Second, we are sensitive to outliers and very non-Gaussian systematic error
distributions.
We later assess the non-Gaussianity of the visit velocity uncertainties by
looking at the visit velocity residuals away from the orbit samples produced by
\thejoker\ using the updated excess variance distribution (see
\sectionname~\ref{sec:discuss-assumptions}).

\section{Companion catalogs}
\label{sec:catalogs}

Using the most likely hyperparameters derived from the initial posterior
samplings (see \sectionname~\ref{sec:inferjitter}), we update and fix the
excess variance prior distribution parameters---$(\mu_{y}, \sigma_{y}) = (9.50,
1.64)$, in units of $\textrm{m}~\textrm{s}^{-1}$---and re-run \thejoker\ on the
parent sample of \apogee\ \DR\ stars.
This approach of estimating a prior \pdf\ from the data and then fixing the
parameters is typically referred to as ``empirical Bayes''; in this case, it is
an approximation to doing a hierarchical inference of the individual system
orbital parameters and the excess variance distribution hyperparameters
simultaneously.

From this run, 91,096 stars completed and successfully returned \nposterior\
samples using \thejoker\ alone.
The remaining 5,135 stars did not return \nposterior\ samples: 4,744 were
flagged as \emph{needs more prior samples}, 391 as \emph{needs MCMC}.
We then proceed to generate \nposterior\ samples for these two subsets using the
methodology explained above (see \sectionname~\ref{sec:samplings}).
The full catalog of posterior samples for all \apogee\ stars in our parent
sample is available online.\footnote{\url{http://adrian.pw/twoface.html}}

We also run on the null comparison sample (see
\sectionname~\ref{sec:control-sample}) with the same parameters.
From the comparison sample run, 13 ``stars'' are flagged as \emph{needs MCMC},
and 1023 are flagged as ``needs more prior samples'';
For these comparison sample stars, we don't continue sampling with \thejoker\ or
MCMC and only use the ($<$\nposterior) posterior samples returned from the
\thejoker.

\subsection{Stars with companions}
\label{sec:catalog-confident}

We do not expect a sharp transition in inferred orbital parameters between stars
with and without companions: There is a continuum of companion properties.
For example, the companions can be low mass, or at long periods, or at high
inclination, all of which will make the velocity semi-amplitude, $K$, small for
a given system.
Therefore, there is no simple cut on radial velocity data alone that would
select a complete sample of stars with companions.
It is nevertheless possible to define a cut that selects stars with
high-confidence companions.

\begin{figure}[h]
\begin{center}
\includegraphics[width=0.8\textwidth]{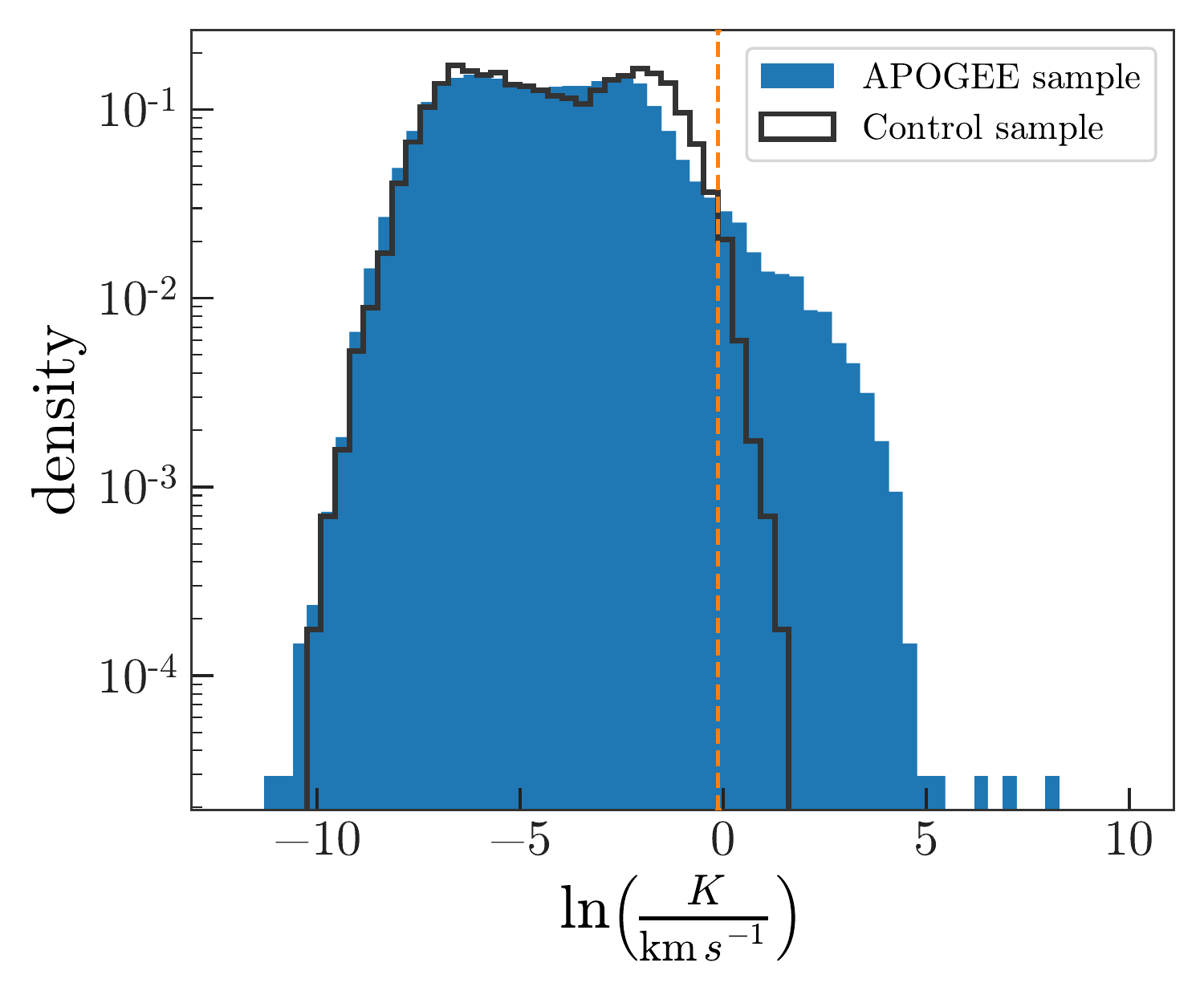}
\end{center}
\caption{%
Distribution of 1st percentiles in $\ln K$ (in units of \kms) for the parent
\apogee\ sample (blue, filled) and for the comparison sample (solid, dark line).
The vertical line (dashed, orange) indicates our adopted cut to select high-$K$
stars that likely have companions at $\ln K = \lnKcut$; $1\%$ of the comparison
sample falls above this cut.
\label{fig:lnK-percentiles}
}
\end{figure}

We select a sample of stars that confidently have companions using percentiles
computed from the log of the posterior samples in the velocity semi-amplitude
parameter, $\ln K$.
\figurename~\ref{fig:lnK-percentiles}, shows the distribution of 1st percentiles
of $\ln K$ computed for all stars with posterior samplings from \thejoker\
(filled, blue histogram), along with the same for posterior samplings for the
null comparison sample (solid, dark line).
To select stars with companions, we use a cut in the 1st percentile of the $\ln
K$ samples such that $1\%$ of the comparison sample is selected, i.e. our
estimated false positive rate is $\approx 1\%$.
We use a threshold of $\ln K = \lnKcut$ to meet this constraint (vertical,
dashed line in \figurename~\ref{fig:lnK-percentiles}); \nhighK\ stars pass this
cut.
For brevity below, we refer to this sample as the ``high-$K$'' stars, and the
complimentary sample as the ``low-$K$'' stars.

\tablename~\ref{tbl:lnK-per} contains a list of all stars in the parent sample
and the value of the 1st percentile over the $\ln K$ samples for each star.
\figurename s~\ref{fig:highK-0} and \ref{fig:highK-1} show eight examples of
high-$K$ stars, i.e. stars that likely have companions, with different numbers
of visits, $N$, indicated on the panels.
\figurename s~\ref{fig:lowK-0} and \ref{fig:lowK-1} shows the same for eight
low-$K$ stars, i.e. stars that either have no companions, low-mass companions,
or companions at long periods or high inclination.
The majority of the stars in the high-$K$ sample have poorly constrained orbital
parameters because the posterior \pdf 's are very multimodal.
The high-$K$ sample selected from \tablename~\ref{tbl:lnK-per} includes all
stars in the catalogs described in the next two subsections, however the
following two catalogs are mutually exclusive.

\subsection{Companions with uniquely-determined orbits}
\label{sec:catalog-unimodal}

Starting from the high-$K$ sample, we select stars that have posterior period
samples that fall within a single mode.
We define a period resolution for each star $\Delta =
\left[4\,P_{\textrm{min}}^2 / (2\pi \, T) \right]$, where $P_{\textrm{min}}$ is
the minimum period sample value, and $T$ is the epoch span of the data for that
star.
We consider a sampling to be unimodal when
\begin{equation}
    P_{\textrm{max}} - P_{\textrm{min}} < \Delta \label{eq:unimodal}
\end{equation}
(see \sectionname~2 of \citealt{Price-Whelan:2017}); \nunimodal\ stars
pass this cut.
Unlike the multi-modal stars, the posterior \pdf 's for these stars can be
approximated using point-estimates and standard deviations of their respective
samplings.
We report maximum \textit{a posteriori} (MAP) sample values for the orbital
parameters, along with other computed quantities for all \nunimodal\ stars in
this high-$K$, unimodal sample.
For stars with measured primary masses, $M_1$, from other work
(\citealt{Ness:2015}), we compute the minimum companion mass, $M_{2,
\textrm{min}}$ and include both masses in this catalog.
We additionally join with the \apogee\ \DR\ \texttt{allStar} catalog and provide
all columns from this catalog for convenience.
\tablename~\ref{tbl:highK-unimodal} contains descriptions and units for all
columns in the high-$K$, unimodal sample catalog, available online.\footnote{\url{http://adrian.pw/twoface.html}}

We have also visually inspected the inferred orbits for all systems in this
sample and have flagged systems with questionable or invalid fits.
The value of the flag is: 0, when the orbits look reasonable, 1, when the orbits
look reasonable but the value of the inferred excess variance is large, and 2,
when the orbits are clearly poor fits.
Many of the stars flagged as ``2'' look like they may be triple systems, as they
tend to have radial velocity variations over two distinct timescales that are
never well-fit by a single Keplerian orbit.
The stars flagged as ``1'' tend to be either upper RGB stars, where
astrophysical surface jitter can lead to RV modulations, or SB2 systems, where
absorption lines from the secondary confuse the RV pipeline and lead to strange
RV signals.
This flag is included in the catalog (\tablename~\ref{tbl:highK-unimodal}) as
the column \texttt{clean\_flag}; To select a clean sample of companions that
have been successfully vetted by-eye, select only stars with
\texttt{clean\_flag == 0}.

\subsection{Companions with highly constrained orbits}
\label{sec:catalog-multimodal}

A subset of the high-$K$ sample (\sectionname~\ref{sec:catalog-confident}) have
multimodal posterior distributions over orbital parameters that are limited to
a few qualitatively different solutions.
For these stars, one or a few more RV measurements would likely lead to
uniquely-determined orbital parameters, making these stars prime candidates for
follow-up efforts.
As examples of such cases, we include an additional catalog of systems that have
effectively bimodal posterior samplings in orbital period.
We identify these systems using $k$-means clustering (\citealt{Lloyd:1982}) of
the posterior samples in period.
In detail, for all stars in the high-$K$ sample, we compute $\ln P$ for all
period samples, then use $k$-means clustering with $k=2$ as implemented in the
\texttt{scikit-learn} package (\citealt{Pedregosa:2011}) to identify two
clusters of samples.
We initialize the cluster positions at either end of the range of possible
periods.
For each cluster, we then ask whether the samples assigned to that cluster are
unimodal (\eqname~\ref{eq:unimodal}).
If the samples in each respective cluster are unimodal, we call the sampling
``bimodal''; \nbimodal\ systems are identified as bimodal.

\figurename~\ref{fig:highK-bimodal} shows a few examples of systems that meet
this criterion.
In many of these cases, there are certain times at which a future observation
would be far more informative.
Visually, those times correspond to regions of time when the predicted RVs have
large variance based on the posterior samples at hand.
For example, in the bottom left panel of \figurename~\ref{fig:highK-bimodal}, an
observation at $t = 60$ would rule out one of the two possible  classes of
orbits.

\tablename~\ref{tbl:highK-bimodal} contains a list of all \apogee\ targets with
bimodal samplings identified in this way.
The table contains two rows for each source, with the period, eccentricity, and
velocity semi-amplitude values from the MAP sample from each period mode.
When available, this table also includes an estimate of the primary mass, $M_1$,
from \citet{Ness:2015} and an estimate of the minimum companion mass, $M_{2,
\textrm{min}}$ for each mode.

We again visually inspect all systems in this sample and assign a flag based on
the apparent quality of the inferred orbits (see
\sectionname~\ref{sec:catalog-unimodal}).
Again, to select a clean sample of companions from this catalog that have been
successfully vetted by-eye, select only stars with \texttt{clean\_flag == 0}.


\section{Results}
\label{sec:results}

Here we highlight a few interesting results from visualizing the properties of
companions in the unimodal and bimodal samples defined above (see \sectionname
s~\ref{sec:catalog-unimodal} and \ref{sec:catalog-multimodal}).
In all figures below, we only plot companions with inferred orbits that pass our
visual inspection (\texttt{clean\_flag == 0}).

\subsection{Systems with unimodal and bimodal posterior samplings}

\begin{figure}[h]
\begin{center}
\includegraphics[width=\textwidth]{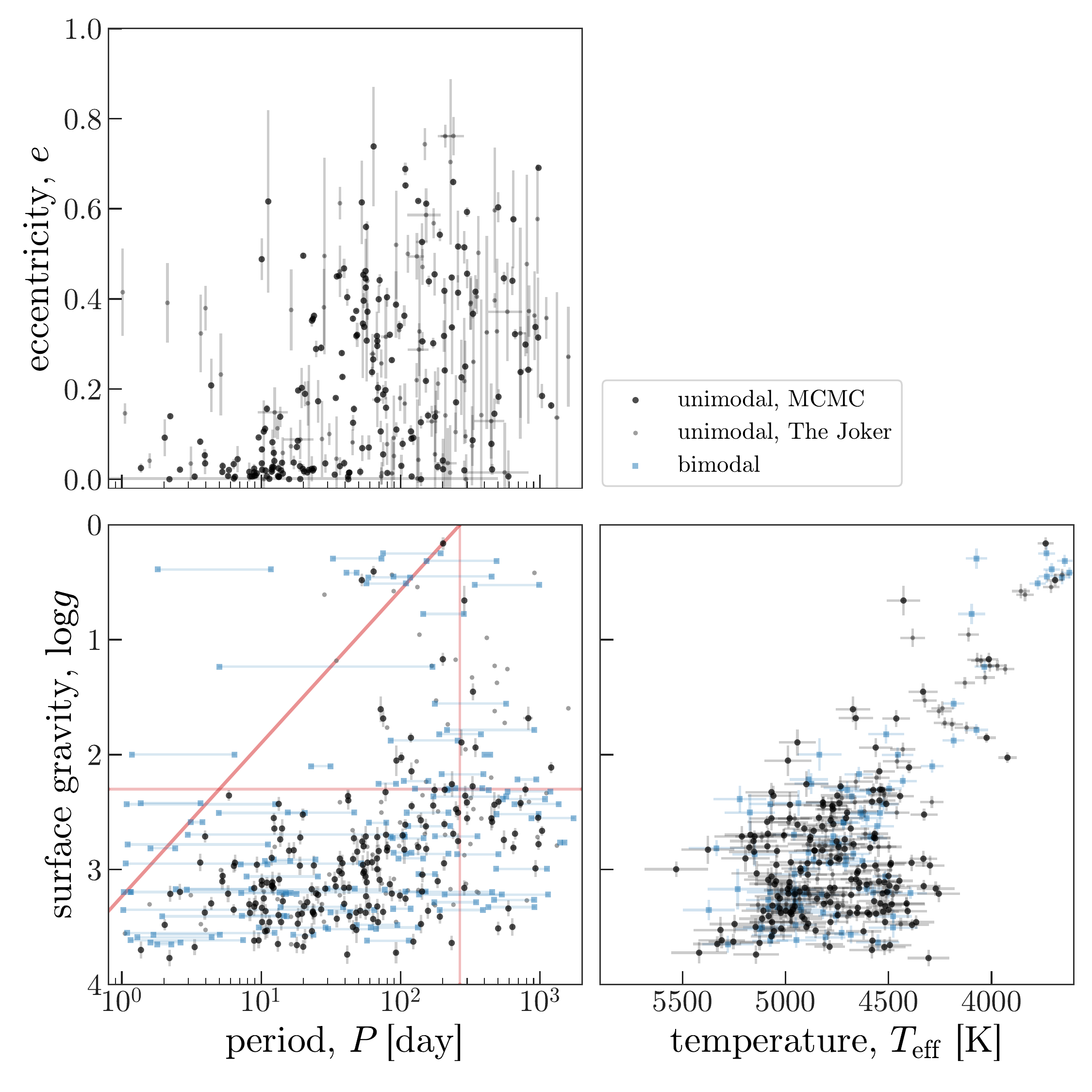}
\end{center}
\caption{%
\textit{Top:} Inferred orbital period and eccentricity for all systems with
unimodal and bimodal posterior samplings from \thejoker.
\textit{Bottom left:} Inferred orbital period and primary star surface gravity,
\logg.
Diagonal line indicates the orbital period of a hypothetical massless companion
at the surface of a $1.35~\msun$ primary star with the given \logg, vertical
line shows the same for $\logg = 0$, and horizontal line shows the rough upper
boundary of the red clump.
\textit{Bottom Right:} Stellar parameters for all primary stars in the unimodal
and bimodal samples.
\label{fig:logg-teff-p-e}
}
\end{figure}

\figurename~\ref{fig:logg-teff-p-e} shows the companion and stellar properties
of the unimodal and bimodal samples: upper left panel shows inferred period and
eccentricity of the binary orbit, lower left panel shows inferred period and
surface gravity of the primary star, lower right panel shows the primary star
stellar parameters.

The period-eccentricity plot (upper left panel in
\figurename~\ref{fig:logg-teff-p-e}) shows a clear signature of tidal
circularization: Such plots for main sequence stars typically show a more
distinct circularization period, below which the majority of stars have
eccentricities close to zero.
Here, we see that close to $P\approx 10~\textrm{day}$ the majority of companion
orbits appear to be circular, but between 10--100~days there appears to be a
gradual decrease in eccentricity rather than a sharp transition.
This is likely because the primary stars in this sample have a large range in
surface gravities and therefore a large range in stellar radii (see lower right
panel).

In the lower left panel of \figurename~\ref{fig:logg-teff-p-e}, the diagonal
(thick) line shows the orbital period of a hypothetical massless companion at
the surface of a $1.35~\msun$ primary star (the median of our sample with
measured masses) with the given surface gravity.
The vertical line in this panel shows the same for a $1.35~\msun$ primary with
$\logg=0$, i.e. the orbital period at the surface of such a star close to the
tip of the RGB.
The horizontal line in this panel shows the approximate upper bound of the red
clump, above which most stars should be fully convective.
Interestingly, there is a dearth of short-period systems for stars above the red
clump ($\logg \lesssim 2.3$) in the triangle defined by the three qualitative
lines, suggestive of possible companion engulfment as the primary star envelope
encases the companion.

The lower right panel of \figurename~\ref{fig:logg-teff-p-e} shows the stellar
parameters of all primary stars in the unimodal and bimodal samples.
Relative to \figurename~\ref{fig:loggteff}, upper RGB stars appear to be
under-represented in these companion catalogs, another tentative signature of
engulfment or depletion of companions on shorter-period orbits.

\subsection{Companion masses and mass ratios}

\begin{figure}[h]
\begin{center}
\includegraphics[width=\textwidth]{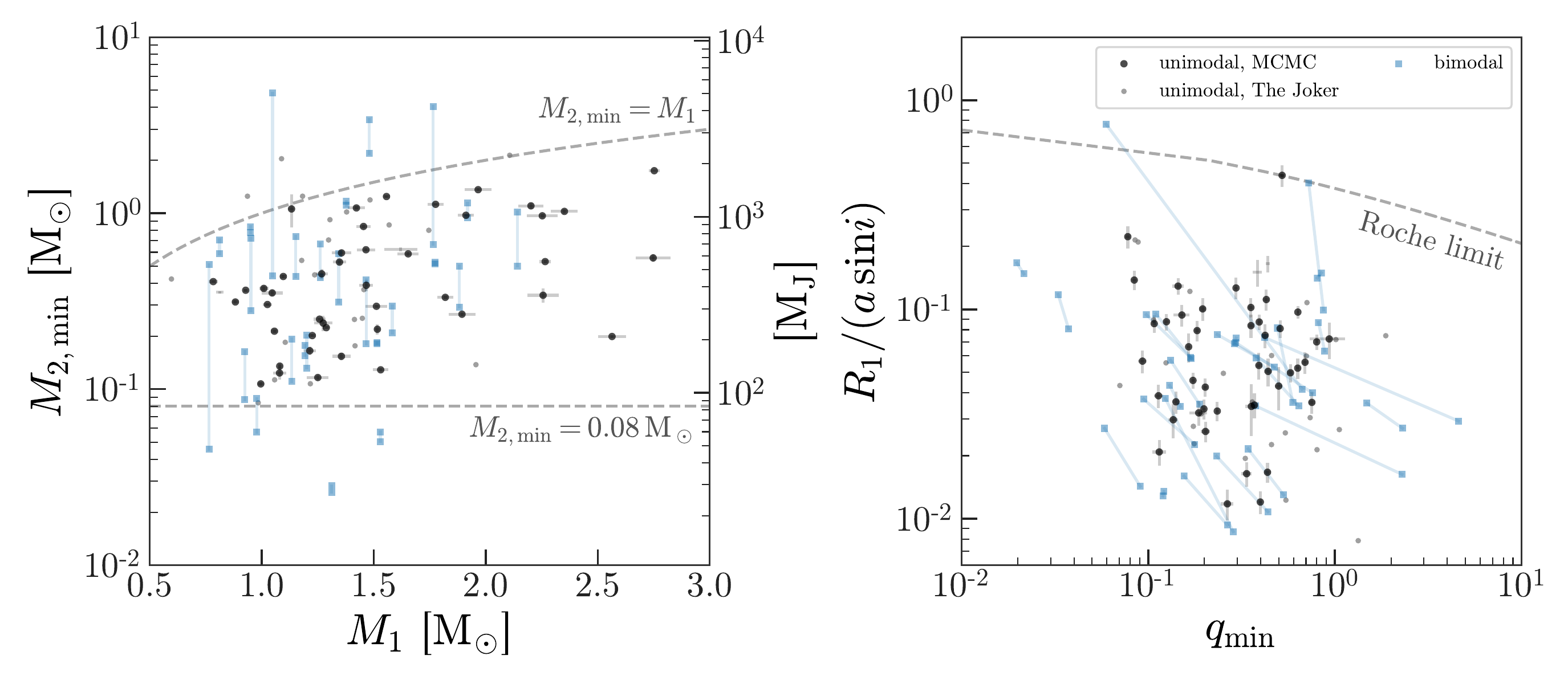}
\end{center}
\caption{%
\textit{Left panel:} Estimated minimum companion mass and primary mass for a
subset of the systems in the high-$K$, unimodal converged sample (31 systems;
black circles), unimodal unconverged sample (38 systems; grey circles), and
bimodal sample (25 systems; blue squares and lines) with $\logg > 2$ and
previously measured primary masses (\citealt{Ness:2015}).
Lines connecting markers (blue) show estimates from each period mode for each
system with bimodal period samplings.
Upper, dashed line shows the equal-mass curve, and lower dashed line shows the
Hydrogen-burning limit.
\textit{Right panel:} Ratio of primary stellar radius, $R_1$, to projected
separation of the two bodies, $a\,\sin i$, and minimum mass ratio,
$q_{\textrm{min}}$, for the same systems and markers in left panel.
\label{fig:mass}
}
\end{figure}

Most of the companions we find have minimum masses $M_{2, \textrm{min}} <
1~\msun$.
\figurename~\ref{fig:mass}, left panel shows primary and companion mass
estimates for systems with unimodal (circle markers) and bimodal (square markes,
blue lines) period samplings for the subset of 69/\nunimodal\ unimodal and
25/\nbimodal\ bimodal systems with primary mass estimates, again using primary
masses from \citet{Ness:2015}.
Upper dashed line shows the $M_{2, \textrm{min}} = M_1$ curve, and lower dashed
line shows the Hydrogen-burning limit, $M_{2, \textrm{min}} = 0.08~\msun$.
For the systems with bimodal samplings, we compute the minimum companion mass
for each period mode and connect these estimates with a line.
Here we restrict to stars with $\logg > 2$ to avoid upper RGB stars that may
have surface oscillations that mimic RV modulations from companions.

\figurename~\ref{fig:mass}, right panel shows the ratio of the primary stellar
radius to the projected separation of the two bodies, $R_1 / (a\,\sin i)$, and
the minimum mass ratio, $q_{\textrm{min}} = M_{2, \textrm{min}} / M_1$, for the
same systems.
Again, circle (black, grey) markers indicate systems from the unimodal sample,
and square (blue) markers and lines connect the two estimates for systems with
bimodal period samplings.
Upper dashed line in this panel indicates the Roche radius, computed using an
approximate functional form (\citealt{Eggleton:1983})
\begin{equation}
    \frac{R}{a} = \frac{0.49\,q^{-2/3}}{0.6\,q^{-2/3} +
        \ln\left(1 + q^{-1/3}\right)}
\end{equation}
where $q = M_2 / M_1$.
All of these systems are consistent with being detached binaries, with the
exception of one notable system (discussed below).

In \figurename~\ref{fig:mass}, a few interesting systems are immediately
obvious: From the left panel, two systems have bimodal period samplings in which
both period modes put the minimum companion mass below the Hydrogen-burning
limit (brown dwarf candidates), and two systems have $M_{2, \textrm{min}} > M_1$
(neutron star or black hole candidates).
In the right panel, one system appears right on the limit of being an
interacting binary ($q_{\textrm{min}} \approx 0.5$), but has no significant UV
flux (\acronym{GALEX} DR5; \citealt{Bianchi:2011}).
These systems are prime candidates for spectroscopic follow-up.

\section{Discussion}

\subsection{Impossible companions and the upper RGB}

Paradoxically, there appear to exist short period ($\lesssim 100~\textrm{day}$)
companions at low surface gravities ($\logg \lesssim 1$), obvious in the
lower-left panel of \figurename~\ref{fig:logg-teff-p-e}: These companions would
orbit within the surface of their primary stars.
Each of these systems appears fine from the \apogee\ data quality flags, but
could nonetheless have incorrect stellar parameters.
As a test, we would ideally be able to compare the spectroscopic stellar
parameters to an independent determination from, e.g., asteroseismology.
Unfortunately, none of these short-period, low-\logg\ stars in the unimodal or
bimodal samples appear in the \acronym{APOKASC} (\citealt{Serenelli:2017})
catalog, which provides asteroseismic stellar parameters for a few hundred
\apogee\ dwarf and giant stars.
We therefore instead cross-match all high-$K$ stars with $\logg < 1$ and $99\%$
of their period samples below $100~\textrm{day}$ to the \acronym{APOKASC}
catalog: One star appears in both samples 2M19024490+4419523.
In \apogee\ \DR, this star has $\logg = 0.57$, but asteroseismic $\logg > 4$
(e.g., \texttt{HUBER\_LOGG}), consistent with being an M dwarf.
Many of these systems could therefore be low-mass dwarf stars with incorrect
stellar parameters in \apogee\ \DR.

Another possibility is that \thejoker\ interprets semi-coherent surface
oscillations (from asteroseismic modes) with sparse time coverage as orbital RV
variations (see, e.g., \citealt{Hekker:2008}).
However, for RGB stars with $\logg < 1$, these modes would likely have
frequencies between $\nu_{\textrm{max}} \approx 0.5$--$5~\mu\textrm{Hz}$
(\citealt{Garcia:2018}), corresponding to periods between $P \approx
20$--$2~\textrm{day}$ and amplitudes between $\approx
20$--$200~\textrm{m}~\textrm{s}^{-1}$ (\citealt{Huber:2011,Huber:2017}).
Except for the lowest \logg\ stars, these amplitudes wouldn't pass our cut on
$K$ (\sectionname~\ref{sec:catalog-confident}).

These systems would need precise, long-term photometric follow-up to obtain
asteroseismic parameters to confirm or explain their existence.

\subsection{Assumptions}
\label{sec:discuss-assumptions}

It is important to keep in mind that the companion catalogs and results
presented in this \documentname\ depend on the assumptions laid out in
\sectionname~\ref{sec:fitting}.
We have only considered radial velocity modulations from a single massive
companion---\emph{no multiple companions}; This is a fundamental limitation of
our search and methodology.
However, stars with multiple companions will likely be included in our companion
catalog anyway, only with two-body orbital solutions that either don't fit the
data well or pick out the shortest period orbit.

Related to the above, we assume that all RV variations are
gravitational---\emph{Kepler}.
We see tentative evidence for surface oscillations at the upper giant branch
(see \figurename~\ref{fig:logg-teff-p-e}, lower left), which we presently treat
as excess variance or intrinsic jitter.
Further observations are needed to determine whether these ``systems'' have
incorrect stellar parameters, or indeed show surface oscillations related to
asteroseismic modes.

We assume that one star in each two-body system overwhelmingly dominates the
luminosity and therefore spectrum of each system---\emph{SB1}.
As seen in \figurename~\ref{fig:mass}, there are some companions with minimum
masses comparable to or consistent with being larger than the masses of the
observed star.
These systems are either (a) nearly edge-on MS--RGB binaries (i.e. consistent
with our assumption), (b) SB2 systems where the \apogee\ pipeline failed to flag
the source as having broad lines or a bad fit, or (c) RGB--stellar remnant
systems with a black hole or neutron star companion.
A subset of these systems that look like they fall in category (c) are being
followed-up to obtain further RV measurements to test whether any companions are
black holes.
However, \thejoker\ can and should be extended to support sampling for SB2
systems with just one additional parameter (typically the mass ratio, or ratio
of velocity semi-amplitudes).

\begin{figure}[h]
\begin{center}
\includegraphics[width=0.6\textwidth]{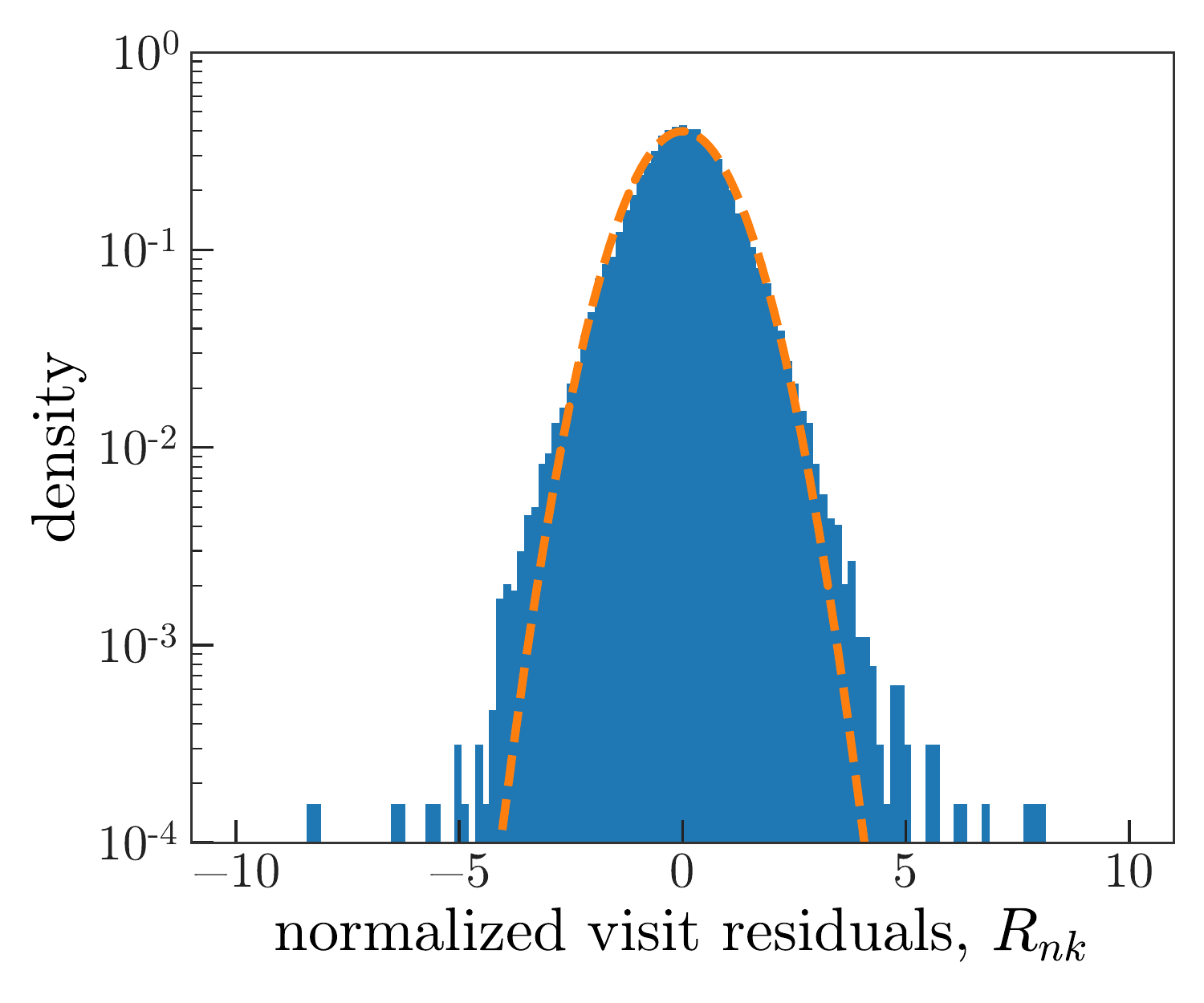}
\end{center}
\caption{%
Histogram shows the distribution of normalized visit residuals (see
\eqname~\ref{eq:normresid}) for all \nvisits\ used in the final run of
\thejoker\ (i.e. with updated excess variance distribution parameters inferred
in \sectionname~\ref{sec:inferjitter}).
The orange curve shows that expected for Gaussian uncertainties.
The distribution appears mostly Gaussian over a large range of residual values,
with slightly more populated tails and evidence of a few catastrophic outliers.
\label{fig:residuals}
}
\end{figure}

Finally, we have assumed that the visit velocity error distribution is Gaussian,
and that the visit velocity uncertainties could be
under-estimated---\emph{simple noise model}.
We can test this assumption using posterior samples from \thejoker\ by computing
the residuals away from our best-fit two-body orbital solutions.
We find that the normalized residuals appear to be very close to Gaussian over a
large dynamic range of normalized residual values, indicating that this
assumption may be sufficient.
\figurename~\ref{fig:residuals} shows the distribution of normalized residuals,
$R_{nk}$, for each visit: the $k$ visit velocities for each $n$ star, $v_{nk}$,
minus the predicted radial velocity from the best-fitting sample returned by
\thejoker, $\hat{v}^*_{nk}$, normalized by the excess-variance-included
uncertainty, $\sigma_{nk}^* = \sqrt{\sigma_{nk}^2 + \hat{s}_{n}^2}$, where
$\hat{s}_{n}$ is computed from the excess-variance parameter of the best-fitting
sample:
\begin{equation}
    R_{nk} = \frac{v_{nk} - \hat{v}^*_{nk}}{\sqrt{\sigma_{nk}^2 +
    \hat{s}_{n}^2}} \quad . \label{eq:normresid}
\end{equation}

\subsection{Comparison to other APOGEE companion catalogs}
\label{sec:compare-troup}

There are at least two other recent catalogs of stellar systems and companions
based on \apogee\ data.

One of these studies focused on decomposing spectra of MS stars into mixtures of
stellar spectra (\citealt{El-Badry:2018}).
Conceptually, this method works because (a) for two unequal-mass stars,
unexpected absorption lines will appear superimposed on the brighter star's
spectrum, and (b) for close to equal-mass stars, the line depths and ratios will
not be well-matched by a single stellar model.
Using this technique, they identified thousands of candidate MS binaries and
trinaries, but did not consider giant stars ($\logg > 4$).
This sample is therefore complimentary to and non-overlapping with the catalog
presented in this work.

The other recent catalog searched for stellar and substellar companions to all
stars in \apogee\ \DRtw\ (including the RGB) that passed a series of quality
cuts, and had $\geq 8$ visits (\citealt{Troup:2016}).
For each star in the sample, orbits were fit to the visit RVs using a multi-step
orbit-fitting procedure: it starts by identifying significant periods and a few
harmonics of those periods, then fits a Keplerian orbit at each of these
harmonics using least-squares fitting (\citealt{De-Lee:2013}) with a modified
$\chi^2$ statistic that penalizes fits in which the phase coverage of the data
is poor.
This procedure is not guaranteed to provide a unique orbit solution.

Of the 382 companions released as a part of the \citet{Troup:2016} search, only
188 of the host stars passed the stellar parameter and quality cuts used to
define the parent sample in this work (see \sectionname~\ref{sec:data}).
We have looked at all of the overlapping stars to compare the previously derived
companion orbital properties to the posterior samplings derived with \thejoker.
We find that the comparisons fall in three categories:
(1) the parameters reported in \citet{Troup:2016} agree with the posterior
samplings, and the period distribution appears unimodal,
(2) the parameters reported in \citet{Troup:2016} identify one possible mode of
a likely multi-modal posterior \pdf\ over orbital parameters, and
(3) the radial velocity data used in \citet{Troup:2016} changed significantly
between \apogee\ \DRtw\ and \DR, so no meaningful comparisons can be made;
roughly 1/3 of the comparison sample falls into each class.
\figurename~\ref{fig:troup-unimodal} shows a few representative cases in which
the \citet{Troup:2016} orbital parameters (orbit shown as orange line in left
panels, parameters shown as orange + in right panels) is consistent with the
orbit samples from \thejoker.
\figurename~\ref{fig:troup-multimodal} shows a few representative cases in which
we find that the posterior \pdf\ over orbital parameters is multimodal, and the
\citet{Troup:2016} orbit identifies one of these modes.
For completeness, \figurename~\ref{fig:troup-datachanged} shows two instances in
which the orbital parameters from \citet{Troup:2016} no longer make sense,
likely because the data changed between data releases.

\subsection{Population inference}

The main motivation for this work was to produce posterior samplings for all
\apogee\ \DR\ stars to be used in a population inference.
To use these samplings for population or hierarchical inference, the orbits of
each individual system don't have to be unimodal and thus \emph{all} \nstars\
samplings can be used in conjunction without well-determined orbital parameters
for the majority of the stars.
These samplings will be useful for constraining (through hierarchical inference)
the period and eccentricity distributions of evolved stars, and the occurrence
rates of companions to evolved stars to test for signatures of engulfment.

\section{Conclusions}

We have selected a catalog of nearly 5,000 stellar systems with companions by
making cuts on posterior beliefs about the amplitude of orbital radial velocity
variations for often low-epoch, sparsely sampled radial velocity data.
We provide posterior samplings over orbital parameters for all \nstars\ in the
parent sample of \apogee\ \DR\ systems used in this work, along with several
sub-catalogs with better constrained orbital information:
\begin{itemize}
    \item A catalog of \nunimodal\ systems with unimodal posterior samplings,
    and therefore uniquely determined orbits, of which 225 are newly discovered
    binary star systems.
    \item A catalog of \nbimodal\ systems with highly constrained posterior
    samplings that, in period, span two distinct period modes.
    For these systems, one or a few more radial velocity measurements would
    uniquely determine their orbits.
    90 of these systems are newly discovered binary star systems.
    \item A catalog of \nhighK\ systems with radial velocity variations
    consistent with having a companion, but which need further radial velocity
    measurements to better constrain the orbital properties.
\end{itemize}
All companion catalogs described in this work (\sectionname~\ref{sec:catalogs})
are available online.\footnote{\url{http://adrian.pw/twoface.html}}

The source code for this project is open source and available from
\url{https://github.com/adrn/TwoFace} under the MIT open source software
license.

\acknowledgements

It is a pleasure to thank
Kevin Covey (WWU),
Marla Geha (Yale),
Marina Kounkel (WWU),
Keith Hawkins (Columbia),
Melissa Ness (Columbia/Flatiron),
David Spergel (Princeton/Flatiron),
and Yuan-Sen Ting (Princeton).

DWH was partially supported by the NSF (grant AST-1517237).
HWR acknowledges support by Sonderforschungsbereich SFB 881 (A3) of the German
Research Foundation (DFG), and from the European Research Council under the
European Union’s Seventh Framework Programme (FP 7) ERC Grant Agreement
n. [321035].
PL was partly funded by "Programa de Iniciaci\'on en Investigaci\'on –
Universidad de Antofagasta."
DAGH and OZ acknowledge support provided by the Spanish Ministry of Economy and
Competitiveness (MINECO) under grant AYA-2017-88254-P.

Funding for the Sloan Digital Sky Survey IV has been provided by the Alfred P. Sloan Foundation, the U.S. Department of Energy Office of Science, and the Participating Institutions. SDSS-IV acknowledges
support and resources from the Center for High-Performance Computing at
the University of Utah. The SDSS web site is www.sdss.org.

SDSS-IV is managed by the Astrophysical Research Consortium for the
Participating Institutions of the SDSS Collaboration including the
Brazilian Participation Group, the Carnegie Institution for Science,
Carnegie Mellon University, the Chilean Participation Group, the French Participation Group, Harvard-Smithsonian Center for Astrophysics,
Instituto de Astrof\'isica de Canarias, The Johns Hopkins University,
Kavli Institute for the Physics and Mathematics of the Universe (IPMU) /
University of Tokyo, Lawrence Berkeley National Laboratory,
Leibniz Institut f\"ur Astrophysik Potsdam (AIP),
Max-Planck-Institut f\"ur Astronomie (MPIA Heidelberg),
Max-Planck-Institut f\"ur Astrophysik (MPA Garching),
Max-Planck-Institut f\"ur Extraterrestrische Physik (MPE),
National Astronomical Observatories of China, New Mexico State University,
New York University, University of Notre Dame,
Observat\'ario Nacional / MCTI, The Ohio State University,
Pennsylvania State University, Shanghai Astronomical Observatory,
United Kingdom Participation Group,
Universidad Nacional Aut\'onoma de M\'exico, University of Arizona,
University of Colorado Boulder, University of Oxford, University of Portsmouth,
University of Utah, University of Virginia, University of Washington, University of Wisconsin,
Vanderbilt University, and Yale University.

The authors are pleased to acknowledge that the work reported on in this
paper was substantially performed at the TIGRESS high performance computer
center at Princeton University which is jointly supported by the Princeton
Institute for Computational Science and Engineering and the Princeton
University Office of Information Technology's Research Computing department.

\software{
    \package{Astropy} (\citealt{Astropy-Collaboration:2013}),
    \package{emcee} (\citealt{Foreman-Mackey:2013}),
    \package{IPython} (\citealt{Perez:2007}),
    \package{matplotlib} (\citealt{Hunter:2007}),
    \package{numpy} (\citealt{Van-der-Walt:2011}),
    \package{scikit-learn} (\citealt{Pedregosa:2011}),
    \package{scipy} (\url{https://www.scipy.org/}),
    \package{schwimmbad} (\citealt{Price-Whelan:2017a}),
    \package{sqlalchemy} (\url{https://www.sqlalchemy.org/}),
    \package{thejoker} (\citealt{Price-Whelan:2017b}).
}

\facility{\sdssiv, \apogee}

\bibliographystyle{aasjournal}
\bibliography{refs}

\clearpage

\appendix
\section{Hierarchical inference of the excess variance parameter}
\label{sec:hierarch}

For each $n$ of $N$ RGB stars in APOGEE, we obtain $M$ posterior samples over
primary orbital parameters $\bs{\theta} = (P, e, \omega, M_0, K, v_0)$ and the
excess-variance parameter, $y = \ln s^2$, using \thejoker; For brevity in
expressions below, we will use the vector
\begin{equation}
    \bs{w} = (\bs{\theta}, y)
\end{equation}
to represent the full set of parameters.
To obtain this sampling, we use an interim (Gaussian) prior on the
excess-variance parameter parametrized by a mean and standard deviation, i.e.
$\alpha_0 = (\mu_{y,0}, \sigma_{y,0})$ as described above.
For a given source, the posterior samples in the above parameters are drawn from
the distribution
\begin{equation}
    \bs{w}_m \sim p(\bs{w}_m \given D, \bs{\alpha}_0)
\end{equation}
where $D$ represents the data for a given object.

We want to compute the likelihood of all data from all $N$ stars, $\{D_n\}$,
given a new set of hyperparameters $\bs{\alpha}$
\begin{equation}
    p(\{D_n\} \given \bs{\alpha}) = \prod_n^N p(D_n \given \bs{\alpha})
\end{equation}
where in the above, we have assumed that this likelihood is separable ( the data
for each source are independent).
The per-source marginal likelihood in the above expression is given by
\begin{align}
    p(D_n \given \bs{\alpha}) &= \int \dd \bs{w}_n \, p(D_n \given \bs{w}_n) \,
      p(\bs{w}_n \given \bs{\alpha})\\
    &= \int \dd \bs{w}_n \, p(D_n \given \bs{w}_n) \, p(\bs{w}_n \given \bs{\alpha}) \,
      \frac{p(\bs{w}_n \given D_n, \bs{\alpha}_0)}{p(\bs{w}_n \given D_n, \bs{\alpha}_0)}\\
    &= p(D_n \given \bs{\alpha}_0) \, \int \dd \bs{w}_n \,
      \frac{p(\bs{w}_n \given \bs{\alpha})}{p(\bs{w}_n \given \bs{\alpha}_0)} \,
      p(\bs{w}_n \given D_n, \bs{\alpha}_0) \label{eq:marglike} \quad .
\end{align}
Using the Monte Carlo integration approximation, \eqname~\ref{eq:marglike} can
be simplified to a sum over prior value ratios of the $M$ posterior samples in
the log-excess-variance parameter for each $n$ star
\begin{equation}
    \approx \frac{\mathcal{Z}_n}{M} \,
      \sum_m^{M} \frac{p(y_{nm} \given \bs{\alpha})}{p(y_{nm} \given \bs{\alpha}_0)}
\end{equation}
where we have canceled the other priors (over $\bs{\theta}$), and all
normalization constants appear in the constant scale factor $\mathcal{Z}_n$.

The above expresion gives the marginal likelihood of the velocity data for a single source given new hyperparameters $\bs{\alpha}$.
The full marginal likelihood is then the product of these individual likelihoods
\begin{align}
    p(\{D_n\} \given \alpha) &\propto \prod_n^N \frac{1}{M} \,
      \sum_m^{M} \frac{p(y_{nm} \given \alpha)}{p(y_{nm} \given \alpha_0)}
      \quad .
\end{align}
In practice, we evaluate the log-marginal-likelihood
\begin{align}
    \ln p(\{D_n\} \given \alpha) &\propto \sum_n^N \left[
      \ln\left( \sum_m^{M} \frac{p(y_{nm} \given \alpha)}{p(y_{nm} \given \alpha_0)} \right)
      - \ln M\right]\label{eq:lnlike}\\
    &\propto \sum_n^N \left[
      \underset{k}{\textrm{logsumexp}}\left[ \ln{p(y_{nm} \given \alpha)} - \ln{p(y_{nm} \given \alpha_0)} \right]
      - \ln M\right]
\end{align}
where $\textrm{logsumexp}$ (the log-sum-exp trick) provides a more stable
estimate of the sum in \eqname~\ref{eq:lnlike}.

\section{Data products}
\label{sec:code-demo}


\begin{table}[ht]
    \centering
    \begin{tabular}{c | c}
    \multicolumn{2}{c}{\textbf{Parent sample of \apogee\ \DR\ stars}}\\
    \hline
    \texttt{APOGEE\_ID} & \texttt{lnK\_per\_1} \\
    \hline
    2M00000002+7417074 & -2.028 \\
    2M00000068+5710233 & -2.600 \\
    2M00000222+5625359 &  2.134 \\
    2M00000446+5854329 & -7.199 \\
    ... & ... \\
    \hline
    \multicolumn{2}{c}{\textit{(\nstars\ rows)}}
    \end{tabular}
    \caption{This table contains \texttt{APOGEE\_ID}'s for all \nstars\ stars
    in the parent sample used in this work, for which we have posterior
    samplings in orbital parameters.
    The other column contains the 1st percentile computed over the $\ln K$
    samples for each source, \texttt{lnK\_per\_1}:
    The ``high-$K$'' sample (see \sectionname~\ref{sec:catalog-confident}) is
    defined using this column, by selecting \texttt{lnK\_per\_1 > 0}.
    }
    \label{tbl:lnK-per}
\end{table}

\begin{table}[ht]
    \footnotesize
    \centering
    \begin{tabular}{l|l|l}
        \multicolumn{3}{c}{\textbf{High-$K$, unimodal systems}} \\
        \hline
        Column name & Unit / format & Description \\
        \hline
        \texttt{APOGEE\_ID}        &                          &
            identifier used by \apogee \\
        \texttt{P}                 & $\mathrm{d}$             & $P$, period \\
        \texttt{P\_err}            & $\mathrm{d}$             & \\
        \texttt{M0}                & $\mathrm{rad}$           &
            $M_0$, phase at reference epoch \\
        \texttt{M0\_err}           & $\mathrm{rad}$           & \\
        \texttt{e}                 &                          &
            $e$, eccentricity \\
        \texttt{e\_err}            &                          & \\
        \texttt{omega}             & $\mathrm{rad}$           &
            $\omega$, argument of pericenter \\
        \texttt{omega\_err}        & $\mathrm{rad}$           & \\
        \texttt{jitter}            & $\mathrm{km\,s^{-1}}$    &
            $s$, excess variance parameter \\
        \texttt{jitter\_err}       & $\mathrm{km\,s^{-1}}$    & \\
        \texttt{K}                 & $\mathrm{km\,s^{-1}}$    &
            $K$, velocity semi-amplitude \\
        \texttt{K\_err}            & $\mathrm{km\,s^{-1}}$    & \\
        \texttt{v0}                & $\mathrm{km\,s^{-1}}$    &
            $v_0$, systemic velocity \\
        \texttt{v0\_err}           & $\mathrm{km\,s^{-1}}$    & \\
        \texttt{t0}                & Barycentric MJD          &
            $t_0$, reference epoch \\
        \texttt{converged}  &                          & binary flag
            indicating whether the sampling converged \\
        \texttt{Gelman-Rubin}      &                          &
            Gelman-Rubin statistic for each MCMC parameter \\
        \texttt{M1}                & $\mathrm{M_{\odot}}$     &
            primary mass estimate (\citealt{Ness:2015}) \\
        \texttt{M1\_err}           & $\mathrm{M_{\odot}}$     & \\
        \texttt{M2\_min}           & $\mathrm{M_{\odot}}$     &
            $M_{2, \textrm{min}}$, minimum $M_2$ mass \\
        \texttt{M2\_min\_err}      & $\mathrm{M_{\odot}}$     & \\
        \texttt{clean\_flag}       &                          &
            [0 = good, 1 = suspicious, 2 = bad], score from by-eye vetting \\
        \texttt{q\_min}            &                          &
            minimum mass ratio \\
        \texttt{q\_min\_err}       &                          & \\
        \texttt{R1}                & $\mathrm{R_{\odot}}$     &
            radius of the primary star \\
        \texttt{R1\_err}           & $\mathrm{R_{\odot}}$     & \\
        \texttt{a\_sini}           & $\mathrm{AU}$            &
            projected separation \\
        \texttt{a\_sini\_err}      & $\mathrm{AU}$            & \\
        \texttt{a2\_sini}          & $\mathrm{AU}$            &
            projected semi-major axis of the companion orbit \\
        \texttt{a2\_sini\_err}     & $\mathrm{AU}$            & \\
        \texttt{DR14RC}            & \texttt{True}/\texttt{False} &
            if the star is included in the \apogee\ \DR\ red clump catalog\\
        \texttt{TINGRC}            & \texttt{True}/\texttt{False} &
            if the star is included in red clump catalog of \citet{Ting:2018}\\
        \vdots & & all columns from \citet{Ness:2015} \\
        \vdots & & all columns from \apogee\ \DR\ \texttt{allStar} file \\
        \hline
        \multicolumn{3}{c}{\textit{(\nunimodal\ rows)}}
    \end{tabular}
    \caption{Description of data table containing summary information for all
    stars in the high-$K$, unimodal sample:
    Stars that likely have companions and have well-determined orbital
    parameters.
    All orbital parameter values are from the maximum \textit{a posteriori}
    (MAP) posterior sample (either from \thejoker, or from \package{emcee}).
    All columns ending in \texttt{\_err} are estimates of the standard-deviation
    of the posterior samples, $\sigma$, computed using the median absolute
    deviation, $\textrm{MAD}$, as $\sigma \approx 1.5 \times \textrm{MAD}$.
    }
    \label{tbl:highK-unimodal}
\end{table}

\begin{table}[ht]
    \centering
    \begin{tabular}{c | c | c | c | c | c | c}
    \multicolumn{7}{c}{\textbf{High-$K$, bimodal systems}}\\
    \hline
    \texttt{APOGEE\_ID} & \texttt{P} & \texttt{e} & \texttt{K} &
        \texttt{M1} & \texttt{M2\_min} & \texttt{clean\_flag} \\
    & $\mathrm{d}$ & & $\mathrm{km\,s^{-1}}$ &
        $\mathrm{M_{\odot}}$ & $\mathrm{M_{\odot}}$ &  \\
    \hline
    2M21362657-0017579 & 14.639 & 0.1171 & 30.68 & -- & -- & 0 \\
    2M21362657-0017579 & 2.3049 & 0.0026 & 28.69 & -- & -- & 0 \\
    2M03180303-0004215 & 35.553 & 0.1302 & 4.541 & 1.20 & 0.20 & 0 \\
    2M03180303-0004215 & 107.22 & 0.3192 & 8.792 & 1.20 & 0.13 & 0 \\
    ... & ... & ... \\
    \hline
    \multicolumn{6}{c}{\textit{(210 rows)}}
    \end{tabular}
    \caption{This table contains \texttt{APOGEE\_ID}'s and limited orbital
    parameter information for all \nbimodal\ stars identified as having bimodal
    posterior samplings in orbital period.
    Each source is listed twice, with MAP values of period, \texttt{P},
    eccentricity, \texttt{e}, and velocity semi-amplitude, \texttt{K}, from each
    period mode of the posterior sampling for the source.
    When available, this table also includes point-mass estimates of the primary
    mass, \texttt{M1}, from \citet{Ness:2015}, and minimum companion masses,
    \texttt{M2\_min}, computed for each period mode.
    }
    \label{tbl:highK-bimodal}
\end{table}

\clearpage

\begin{figure}[hp]
\begin{center}
\includegraphics[width=\textwidth]{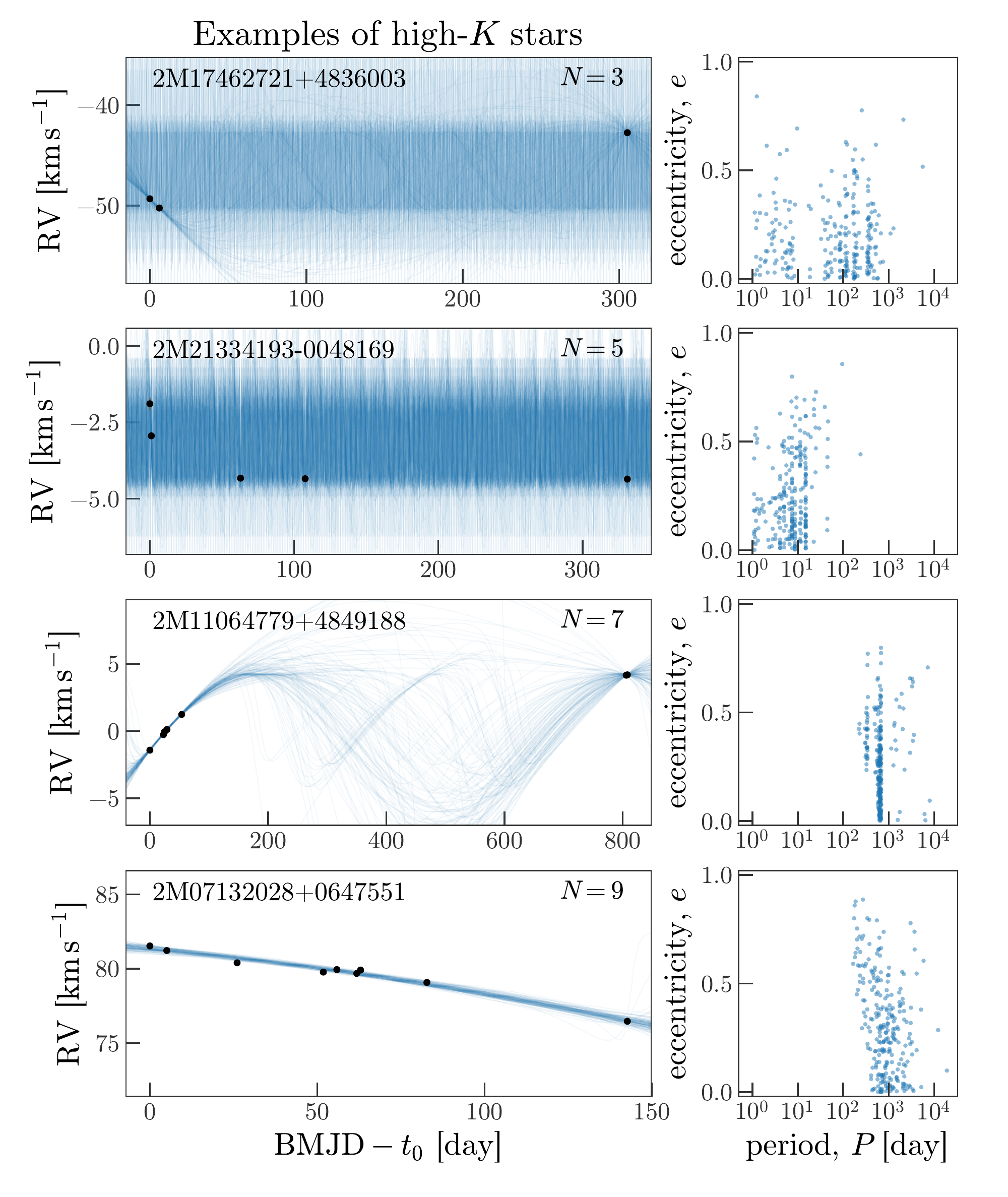}
\end{center}
\caption{%
Examples of stars in our high-$K$ sample, i.e. stars that likely have companions
(see \sectionname~\ref{sec:catalog-confident}), with different numbers of
visits.
Left panels show the data (black markers, error bars are the visit velocity
uncertainties) with 128 random orbits from the \nposterior\ posterior samples
under-plotted (lines, blue); The \texttt{APOGEE\_ID} of each target and the
number of visits that pass our quality cuts, $N$, are indicated on each panel.
Right panels show the \nposterior\ posterior samples visualized in
period--eccentricity space.
In most cases, despite confidently having companions based on the RV amplitude,
the permitted orbit fits are highly multimodal.
\label{fig:highK-0}
}
\end{figure}

\begin{figure}[hp]
\begin{center}
\includegraphics[width=\textwidth]{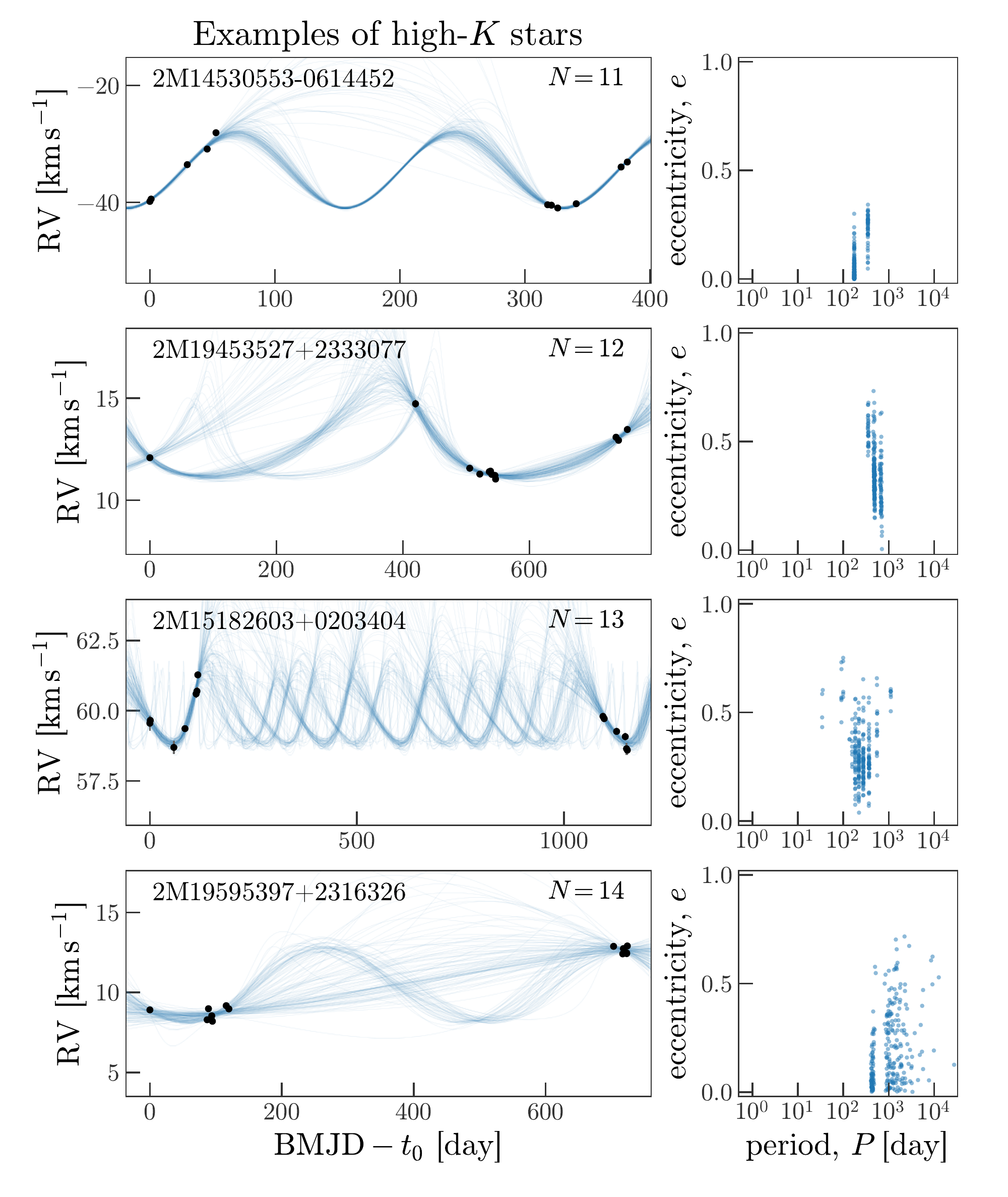}
\end{center}
\caption{%
Continuation of \figurename~\ref{fig:highK-0}.
\label{fig:highK-1}
}
\end{figure}

\begin{figure}[hp]
\begin{center}
\includegraphics[width=\textwidth]{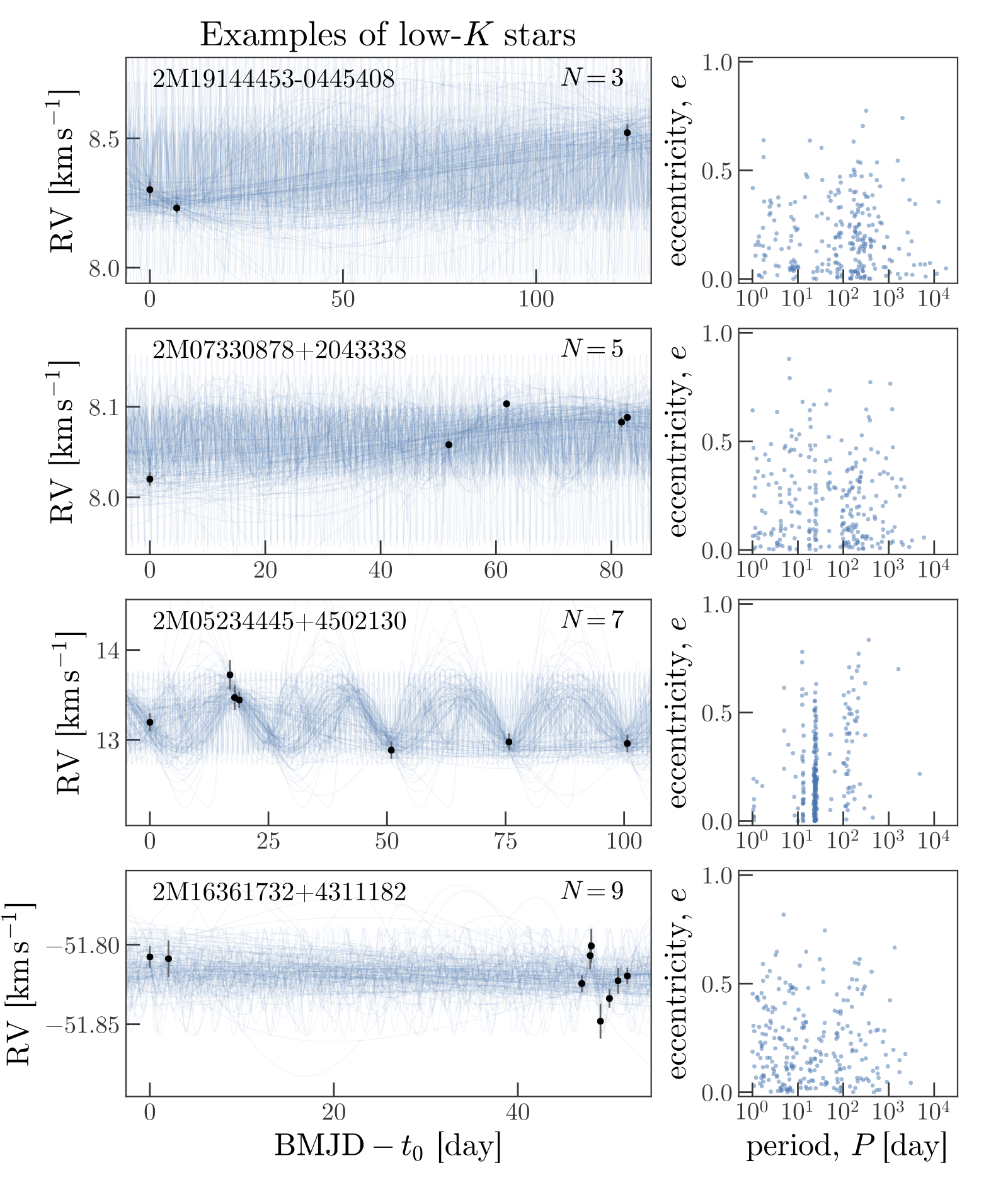}
\end{center}
\caption{%
Same as \figurename~\ref{fig:highK-0}, but for stars in the low-$K$ sample, i.e.
stars that have undetected or no companions.
\label{fig:lowK-0}
}
\end{figure}

\begin{figure}[hp]
\begin{center}
\includegraphics[width=\textwidth]{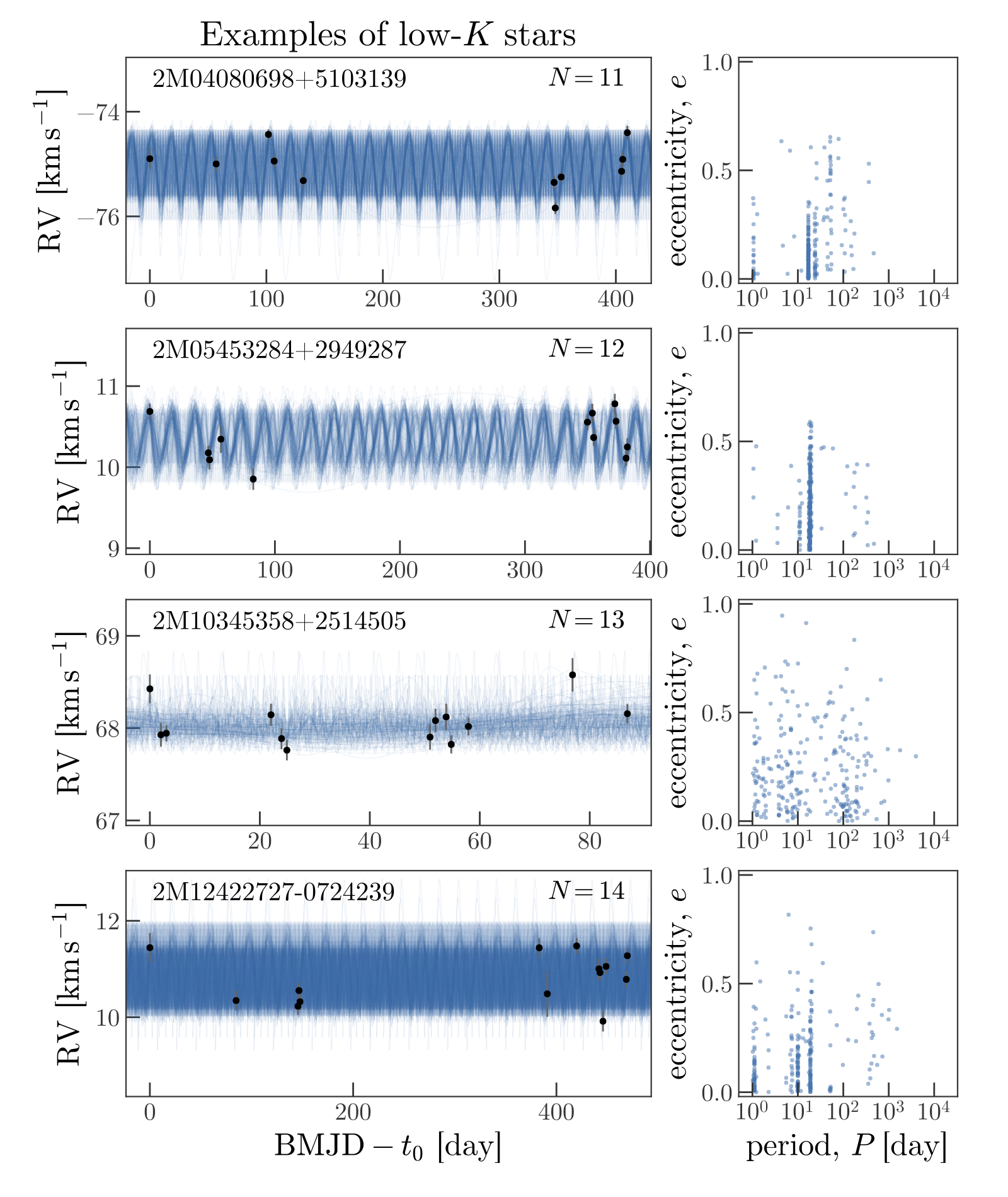}
\end{center}
\caption{%
Continuation of \figurename~\ref{fig:lowK-0}.
\label{fig:lowK-1}
}
\end{figure}

\begin{figure}[hp]
\begin{center}
\includegraphics[width=\textwidth]{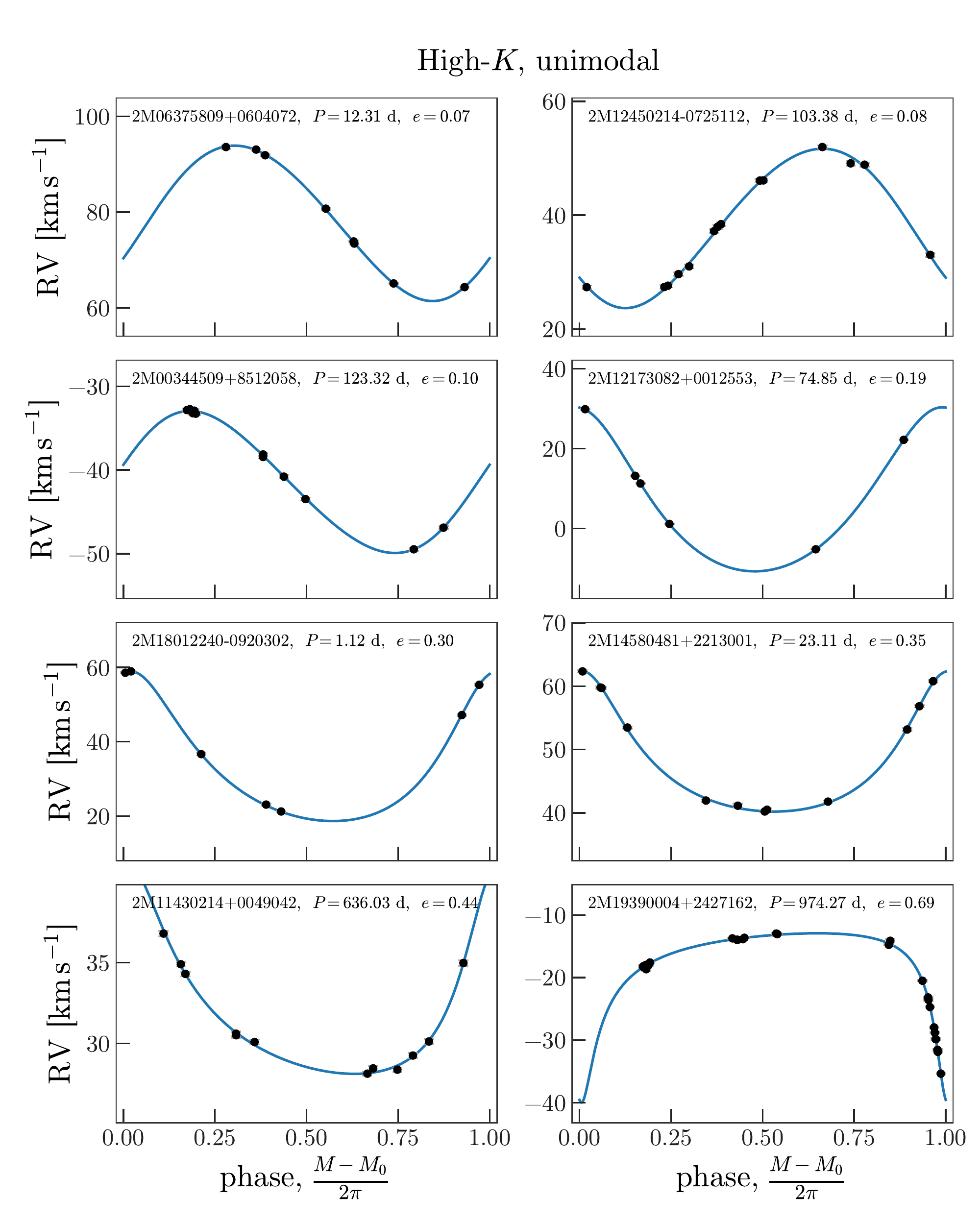}
\end{center}
\caption{%
Randomly-selected examples of high-$K$ stars with effectively unimodal posterior
samplings and converged MCMC samplings.
Each panel shows the data (black markers), phase-folded at the period of the
posterior sample with maximum posterior probability.
Visit velocity uncertainties are shown as black error bars (these are typically
smaller than or equal to the size of the markers), and the inferred jitter as
grey, capped error bars.
Line (blue) shows the orbit compute from the posterior sample with maximum
posterior probability.
\label{fig:highK-unimodal}
}
\end{figure}

\begin{figure}[hp]
\begin{center}
\includegraphics[width=\textwidth]{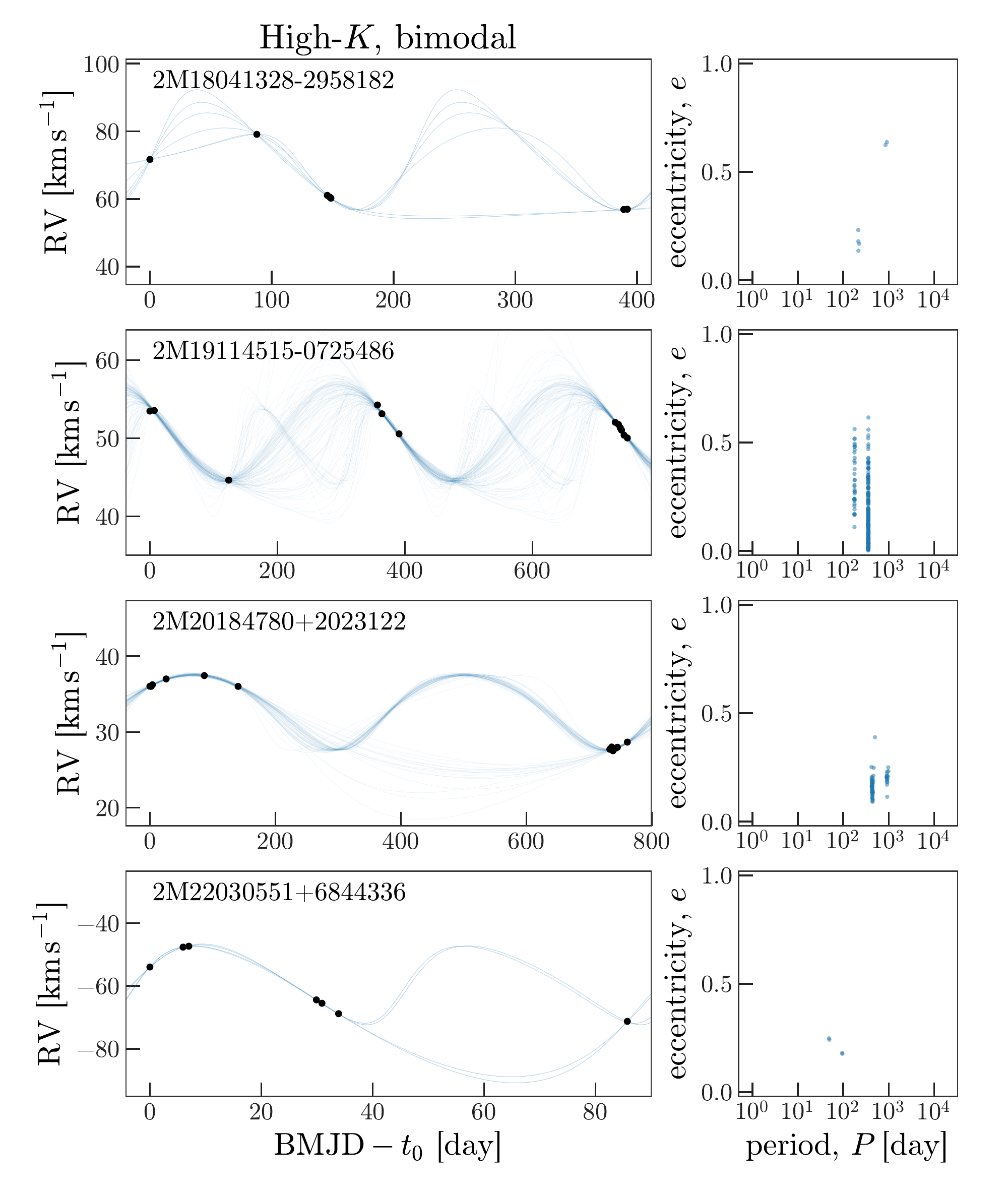}
\end{center}
\caption{%
Randomly-selected examples of high-$K$ stars with bimodal posterior samplings.
Same as \figurename~\ref{fig:highK-unimodal}, but for bimodal posterior
samplings.
\label{fig:highK-bimodal}
}
\end{figure}

\begin{figure}[h]
\begin{center}
\includegraphics[width=\textwidth]{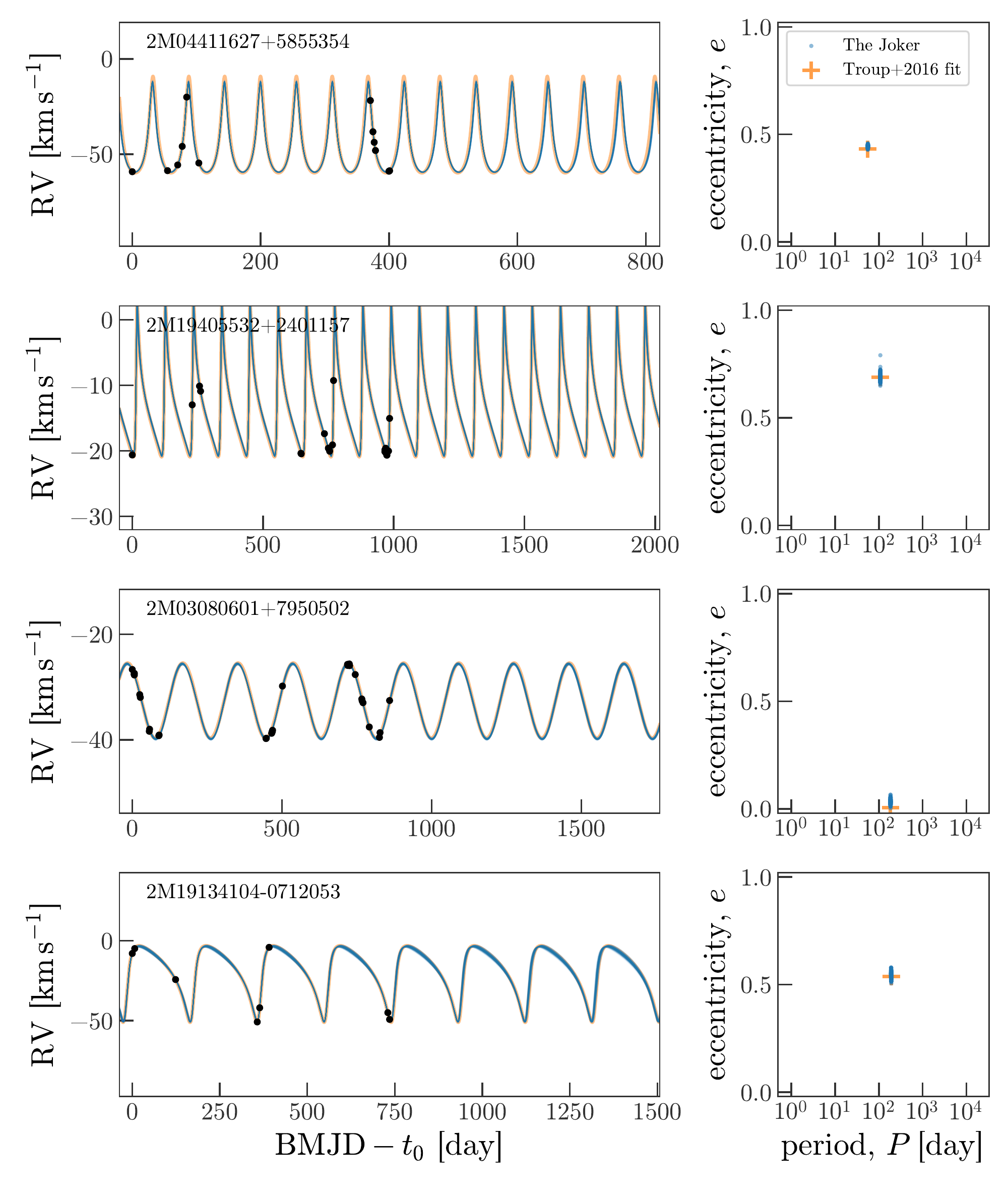}
\end{center}
\caption{%
Four examples of stars that also appear in the \citet{Troup:2016} companion
catalog where our orbit samplings are unimodal, and agree well with the previous
orbit fits.
Left panels show the data from \apogee\ \DR\ (black markers), the orbit fit
(orange, thick line; \citealt{Troup:2016}), and 128 orbits computed from
posterior samples generated in this work (blue, thinner lines).
Right panels show the period and eccentricity of the orbit fit (orange, +
marker; \citealt{Troup:2016}) and the posterior samples from this work (blue
markers).
\label{fig:troup-unimodal}
}
\end{figure}

\begin{figure}[h]
\begin{center}
\includegraphics[width=\textwidth]{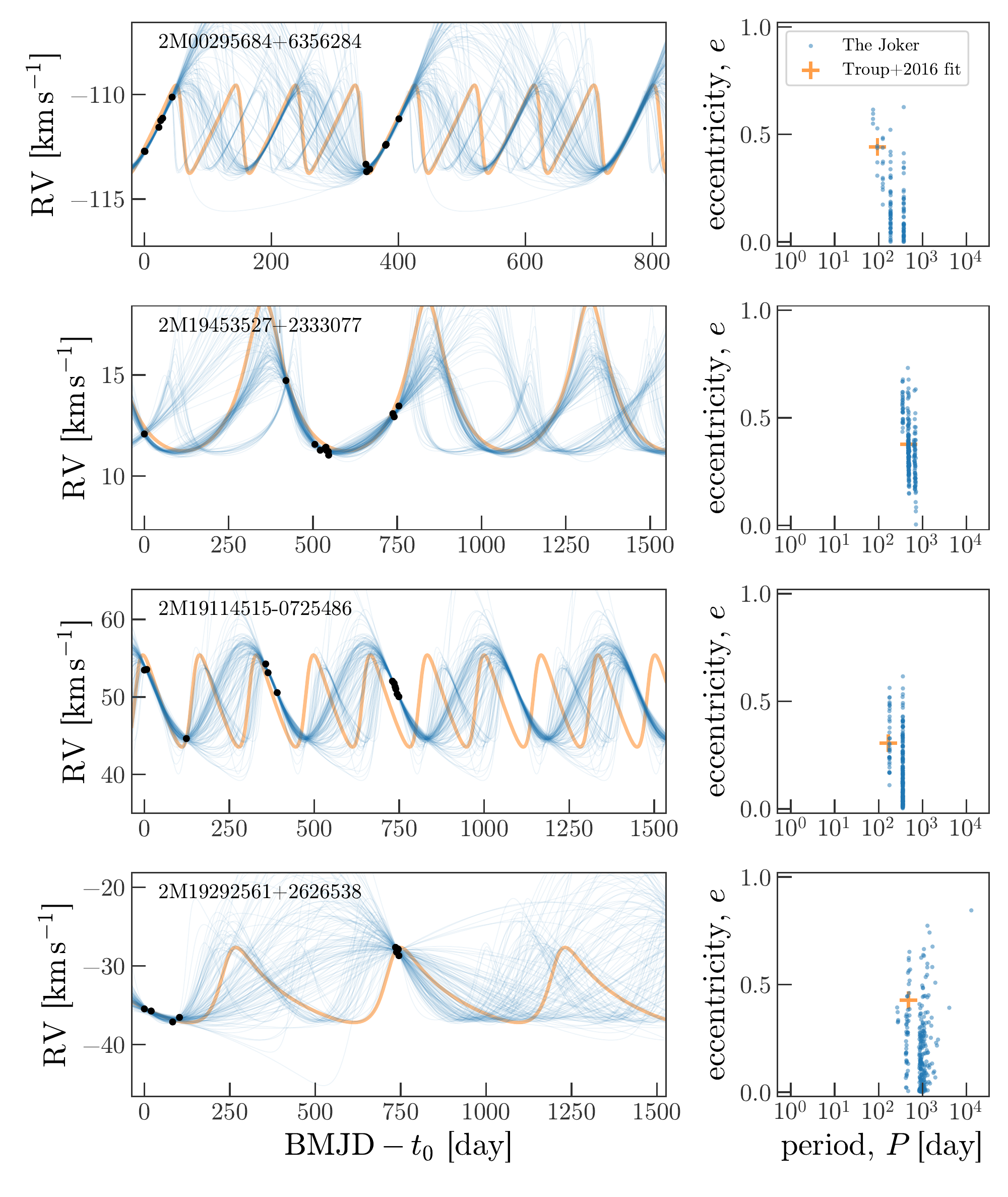}
\end{center}
\caption{%
Same as \figurename~\ref{fig:troup-unimodal}, but for four stars where we find a
multi-modal posterior sampling.
In this cases, the orbit fit from \citet{Troup:2016} generally identify one of
the possible period modes, but multiple orbit solutions are typically allowed.
\label{fig:troup-multimodal}
}
\end{figure}

\begin{figure}[h]
\begin{center}
\includegraphics[width=\textwidth]{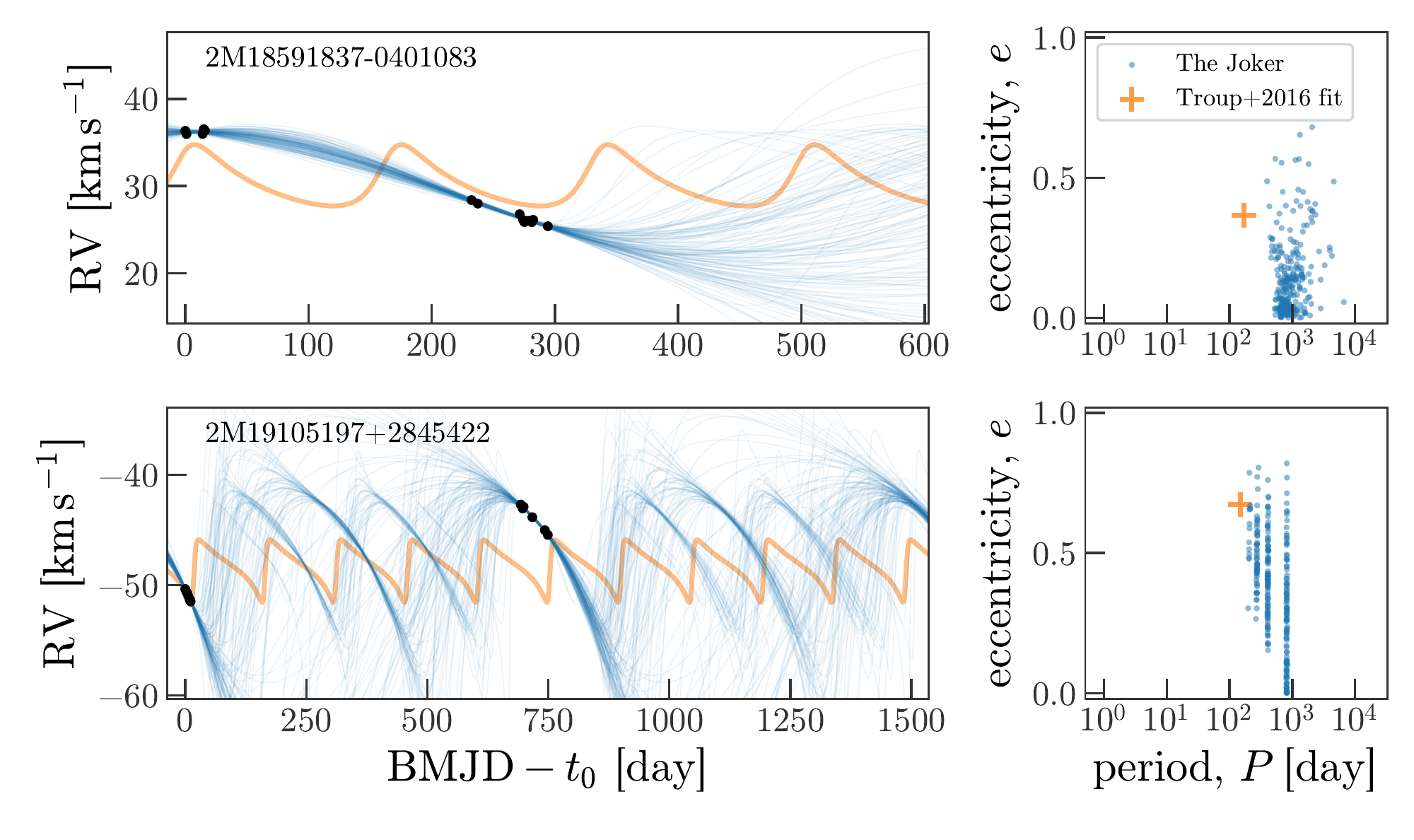}
\end{center}
\caption{%
Same as \figurename~\ref{fig:troup-unimodal} and
\figurename~\ref{fig:troup-multimodal}, but for two stars in which no meaningful
comparison can be made because the visit velocities changed significantly
between \DRtw\ and \DR.
\label{fig:troup-datachanged}
}
\end{figure}

\end{document}